%% file: conference_041818.tex
\RequirePackage[T1]{fontenc}
\documentclass[conference]{IEEEtran}

\usepackage[left=0.63in,right=0.625in,top=0.75in,bottom=1in]{geometry}
\IEEEoverridecommandlockouts
\usepackage{amsmath,amssymb,amsfonts, bm}
\usepackage{algorithmic}
\usepackage{graphicx}
\usepackage{textcomp}
\usepackage{xcolor}
\usepackage{glossaries} 
\usepackage[numbers,sort&compress]{natbib}
\usepackage[font=small]{caption}

\makeglossaries%
\loadglsentries{abbr}

\usepackage{tikz}
\usepackage{pgfplotstable}
\usepgfplotslibrary{groupplots,dateplot}
\usetikzlibrary{patterns,shapes.arrows}
\pgfplotsset{compat=newest}
\usetikzlibrary{spy}
\usepgfplotslibrary{polar}
\usetikzlibrary{positioning}
\usetikzlibrary{plotmarks}
\usetikzlibrary{arrows}
\usetikzlibrary{arrows.meta}
\usepgfplotslibrary{patchplots}
\usetikzlibrary{decorations, decorations.text,backgrounds} 
\usepgfplotslibrary{colorbrewer}
\usetikzlibrary{pgfplots.statistics}

\usepackage{threeparttable}
\usepackage{booktabs}
\usepackage{url}
\usepackage[binary-units=true]{siunitx}
\DeclareSIUnit{\Wh}{Wh}
\DeclareSIUnit{\belmilliwatt}{Bm}
\DeclareSIUnit{\dBm}{\deci\belmilliwatt}
\DeclareSIUnit{\belmillii}{Bi}
\DeclareSIUnit{\dBi}{\deci\belmillii}

\definecolor{color0}{rgb}{0.12156862745098,0.466666666666667,0.705882352941177}
\definecolor{color1}{rgb}{1,0.498039215686275,0.0549019607843137}
\definecolor{color2}{rgb}{0.172549019607843,0.627450980392157,0.172549019607843}

\usepackage{subcaption}
\usepackage{comment}

\newcommand*{\vectr}[1]{\MakeLowercase{\boldsymbol{\mathrm{#1}}}}
\newcommand*{\matr}[1]{\MakeUppercase{\boldsymbol{\mathrm{#1}}}}

\newcommand*{\hermconj}[1]{{#1}^{H}}
\newcommand*{\inC}[1]{\in\mathbb{C}^{#1}}
\newcommand{\norm}[1]{\left\lVert#1\right\rVert}
\newcommand{\abs}[1]{\left\lvert#1\right\rvert}
\newcommand{\expt}[1]{\mathbb{E} \left\{#1\right\}}

\newcommand{\cn}[2]{\ensuremath{\sim\mathcal{C}\mathcal{N}\left(#1,#2\right)}}

\usepackage{xifthen}
\newcommand{\prob}[1][]{
\ifthenelse{\isempty{#1}}%
      {\ensuremath{P}}%
    {\ensuremath{P\left\(#1\right\)}}%
}

\newcommand{\normalized}[1]{\bar{#1}}

\newcommand*{\vect}[1]{\vectr{#1}}
\newcommand*{\mat}[1]{\matr{#1}}

\newcommand{\diag}{\text{diag}}

\newcommand{\francois}[1]{{\color{cyan}[Francois: #1]}}

\pgfplotsset{every axis/.append style={
                    label style={font=\footnotesize},
                    tick label style={font=\footnotesize}  
                    }}

\def\BibTeX{{\rm B\kern-.05em{\sc i\kern-.025em b}\kern-.08em
    T\kern-.1667em\lower.7ex\hbox{E}\kern-.125emX}}
\begin{document}

\title{Grant-Free Random Access in Massive MIMO for Static Low-Power IoT Nodes%
\thanks{The project has received funding from the European Union’s Horizon 2020 research and innovation programme under grant agreement No 101013425.}
}

\author{\IEEEauthorblockN{Gilles Callebaut,
        Liesbet Van der Perre and		
        Fran\c{c}ois Rottenberg
		}
\IEEEauthorblockA{KU Leuven, ESAT-WaveCore, Ghent Technology Campus, 9000 Ghent, Belgium}
\IEEEauthorblockA{E-mail: gilles.callebaut@kuleuven.be}
}

\maketitle
\begin{abstract}
Massive MIMO is a promising technology to enable a massive number of \acrlong{iot} nodes to transmit short and sporadic data bursts at low power. In conventional cellular networks, devices use a grant-based random access scheme to initiate communications. This scheme relies on a limited set of orthogonal preambles, which simplify signal processing operations at network access points. However, it is not well suited for \gls{iot} devices due to: (i)~the large protocol overhead, and (ii)~the high probability of collision. In contrast to the grant-based scheme, a grant-free approach uses user-specific preambles and has a small overhead, at the expense of more complexity at access points. In this work, a grant-free method is proposed, applicable for both co-located and cell-free deployments. The method has a closed form solution, which results in a significantly lower complexity with respect to the state-of-the art. The algorithm exploits the static nature of \gls{iot} devices through the use of prior channel state information. With a power budget of \SI{1}{\milli\watt}, \num{64} antennas are sufficient to support \num{1000} nodes with \num{200} simultaneous access requests with a probability of false alarm and miss detection below $10^{-6}$ and $10^{-4}$, respectively.
\end{abstract}

\begin{IEEEkeywords}
cell-free, grant-free, initial access, internet-of-things, massive mimo, random access
\end{IEEEkeywords}

\section{Introduction}


In \gls{iot} networks, a high number of devices are connected. However, only a handful of these devices are active at the same time, resulting in sporadic uplink transmissions. 
In order to serve these nodes, the active devices need to be detected, prior to decoding the data. Due to this sporadic traffic and the high number of \gls{iot} devices, allocating orthogonal preambles to these devices would incur an unacceptable overhead. As the network is unable to predict when these devices are active, an adequate initial access scheme needs to be implemented. The specificities of massive \gls{iot}, i.e., energy-limited, uplink focused, and low-payload size transmissions, call for a tailored initial access scheme. 

\begin{figure}[btp]
    \centering
    \begin{subfigure}[t]{0.47\linewidth}
        \includegraphics[width=\linewidth]{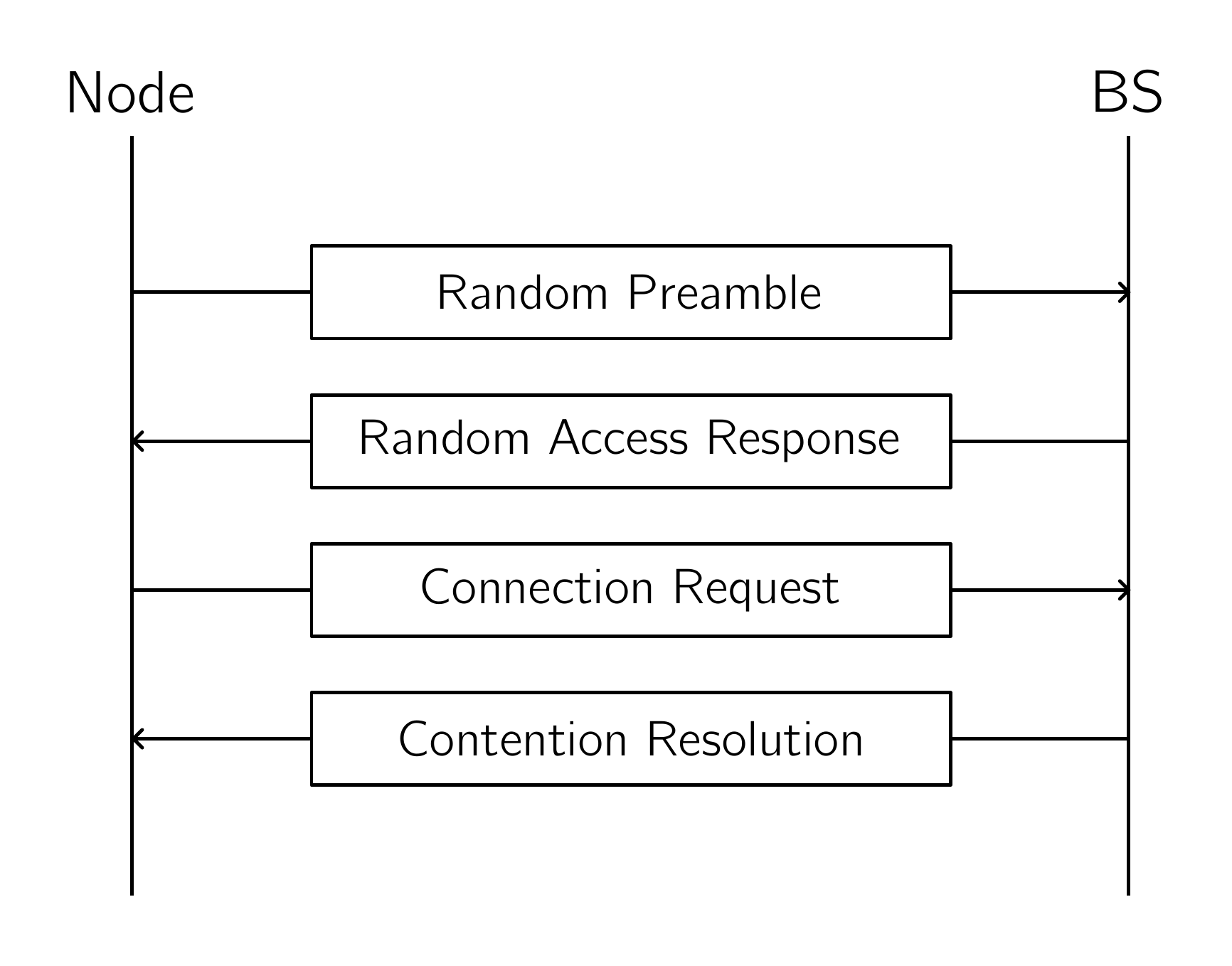}
        \caption{A typical grant-based random access scheme. If the same random preamble is selected the devices are required to resolve their collisions.}\label{fig:grant-based}
    \end{subfigure}\hfill%
    \begin{subfigure}[t]{0.47\linewidth}
        \includegraphics[width=\linewidth]{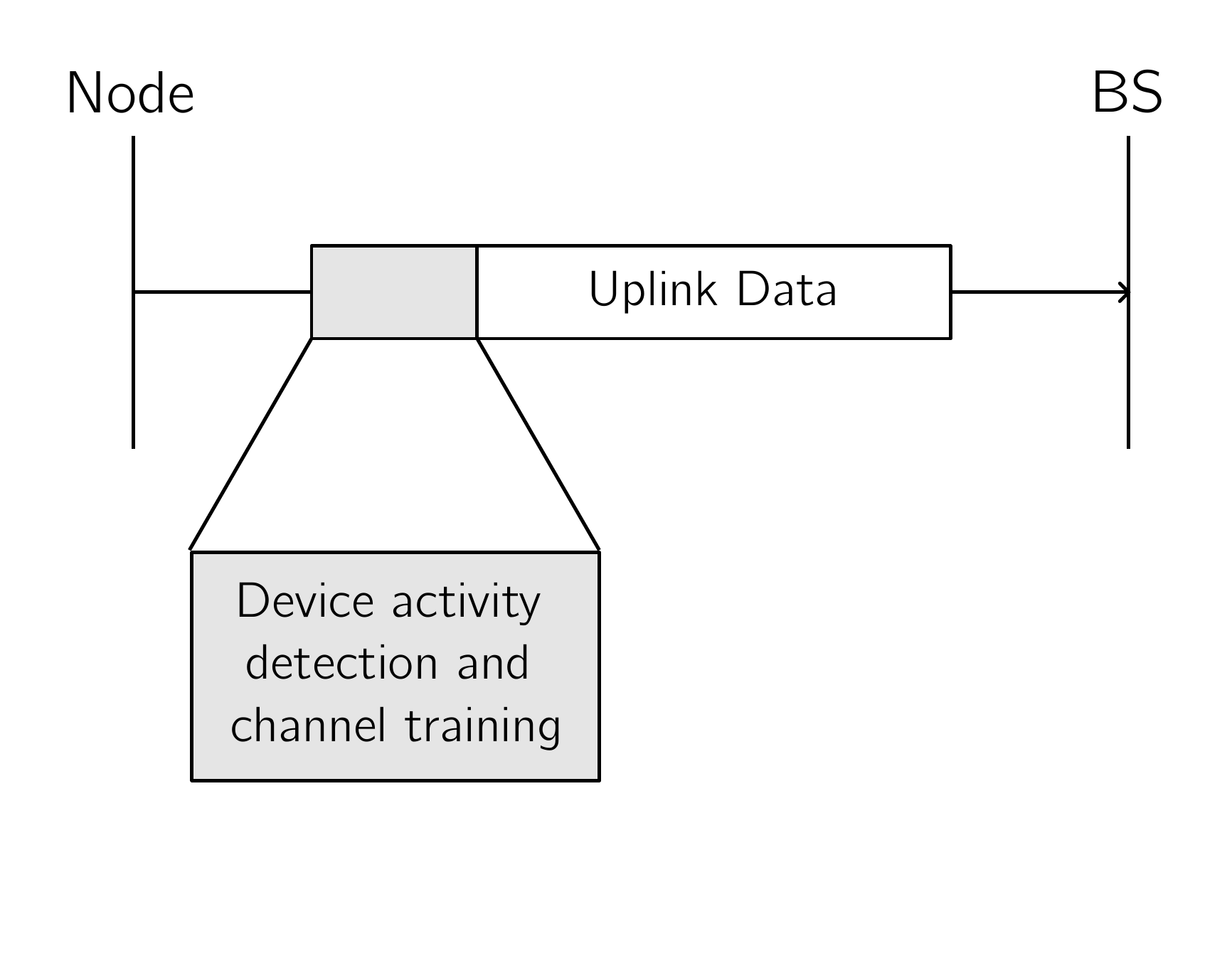}
        \caption{A grant-free random access procedure. Prior to data, a unique pilot sequence is transmitted for device activity detection and channel estimation.}\label{fig:grant-free}
    \end{subfigure}\hfill%
    \caption{Grant-based and grant-free random access mechanisms.}
\end{figure}

Several multiple access techniques have been proposed for massive \gls{iot}~\cite{TUSETPEIRO201584,7390876,6666156,8360103,8734871}. These techniques follow a grant-based or grant-free approach, as illustrated in Fig.~\ref{fig:grant-based} and~\ref{fig:grant-free}, respectively. In the former, the devices request access to the network, prior to the communication. This is commonly done by competing for a dedicated orthogonal pilot sequence. In the latter approach, the devices are not required to request access and other approaches to mitigate and resolve potential collisions are implemented.
An overview of different techniques can be found in literature~\cite{BANA2019100859, 7891796}.

\textbf{Grant-based random access.}
Based on the random access protocol used in \gls{lte}, \citeauthor{bjornson_random_2017-1}~\cite{bjornson_random_2017-1} propose the \gls{sucre} protocol, illustrated in Fig.~\ref{fig:grant-based}. Each active device requests a dedicated and unique orthogonal pilot by transmitting a random access pilot, with the possibility of using the same pilot as another contending device. The base station estimates the channel to each user based on the received pilots, after which it responds with orthogonal precoded downlink pilot signals, corresponding to the used pilots in the uplink. In case multiple devices have used the same pilot, the downlink signal is multicasted in a maximum ratio transmission fashion towards these devices, causing the expected received signal strength to be lower than expected. The (average) expected signal strength can be determined at the device side thanks to the channel hardening effect experienced in massive MIMO. Each device can, hence, detect whether a collision occurred. In this protocol, the strongest user is considered the winner and is scheduled for communication using a dedicated and unique pilots sequence. Other work has extended this technique by introducing, a.o., a higher fairness among the contenders~\cite{8886093,8935438}.

\textbf{Grant-free random access.}
In contrast to the grant-based random access, in a grant-free scheme, the devices use preassigned pilot sequences~\cite{9154288}. As shown in Fig.~\ref{fig:grant-free}, by skipping the grant request and collision resolution, both the pilot and data are sent in a single step. As mentioned earlier, the disadvantage of this approach is that it is not possible to utilize orthogonal pilots because the number of devices is much larger than the preamble length, i.e., $K \gg \tau_p$. The challenge in grant-free random access is device activity detection. 
Taking advantage of the sparse nature of the activity of the devices, different algorithms have been proposed to tackle this challenge through compressed sensing techniques~\cite{8462577,8454392, 8961111, 8683324}. 
These algorithms perform well because of the high number of \gls{ap} antennas in massive MIMO, facilitating device detection in massive \gls{iot} networks.

\citeauthor{FenglerAsilomar}~\cite{FenglerAsilomar,FenglerTIT} study the performance of several estimators to detect the active devices for a co-located massive MIMO system. In~\cite{GanesanSPAWC,GanesanTCOM}, this work is extended to the cell-free case by considering different large-scale fading coefficients per \gls{ap}. They, however, were unable to directly use the proposed algorithm of~\cite{FenglerTIT} and had to consider only the contribution of the most dominant \gls{ap}.



In this work, we consider the generic case where devices are not scheduled, and they transmit whenever needed, i.e., they are uplink-centric. As in conventional systems, we consider that all devices are time synchronized and work in a time slotted fashion. This can be easily achieved through downlink synchronization reference signals. Because of the large number of devices, providing orthogonal preambles to each user would generate a too large pilot overhead, i.e., pilots of length~$K$. 
In this algorithm, a unique but non-orthogonal preamble is randomly generated and assigned to each user. Similar to key distribution in conventional \gls{lpwan}, this sequence is known by both the network and the device, and is easy to implement.
Based on prior \gls{csi}, the active devices are detected. In case both the \gls{iot} node and \glspl{ap} are static, i.e., immobile, the channel response will be static in time as long as the environment is static as well. Hence, we can expect that the previously estimated \gls{csi} can be used for a longer time period than assumed in literature~\cite{marzetta2016fundamentals}, given the static nature of the devices.

The proposed approach has a lower complexity than the techniques reported in literature. Our algorithm is evaluated through numerical simulations. Next to conventional massive MIMO, i.e., co-located massive MIMO, a cell-free deployment is included in the investigation.

\textbf{Notations:} vectors are denoted by boldface lower case~\(\vectr{x}\) and matrices by boldface capital letters~\(\matr{X}\).
The superscript \(\hermconj{(\cdot)}\) is used to denote the conjugate transpose operation. The absolute value is denoted by \(\abs{\cdot} \). The Kronecker product is denoted by $\otimes$.
The notation \(\expt{\cdot}\) denotes the expectation of a random variable. The set of complex numbers is denoted by the symbol \(\mathbb{C}\). The uniform distribution bounded by the closed interval between $a$ and $b$ is denoted by \(\mathcal{U}_{[a, b]}\). The complex normal and normal distribution with mean $\mu$, standard deviation $\sigma$ and covariance matrix $\matr{\Gamma}$ is denoted by $\mathcal{C}\mathcal{N}\left(\mu,\matr{\Gamma}\right)$, respectively. The identity matrix $\matr{I}_n$ is a $n\times n$ square matrix with ones on the main diagonal and zeros elsewhere.

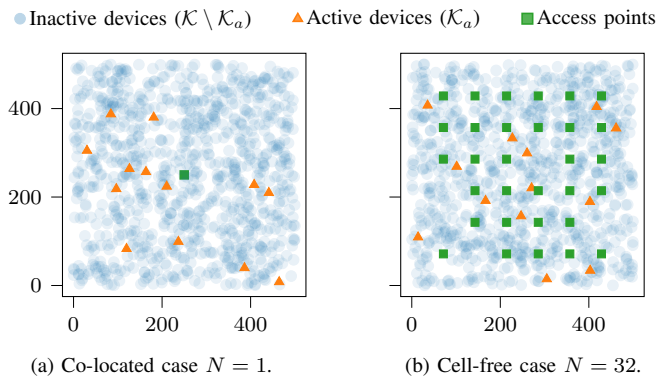
\begin{figure}[btp]
    \centering
    \begin{subfigure}{\linewidth}
        \centering

        \begin{tikzpicture} 
            \definecolor{color0}{rgb}{0.12156862745098,0.466666666666667,0.705882352941177}
            \definecolor{color1}{rgb}{1,0.498039215686275,0.0549019607843137}
            \definecolor{color2}{rgb}{0.172549019607843,0.627450980392157,0.172549019607843}
            \begin{axis}[%
            hide axis,
            xmin=10,
            xmax=50,
            ymin=0,
            ymax=0.4,
            legend style={draw=white!15!black,legend cell align=left},
            legend columns=3,
            legend cell align={left},
            legend style={fill opacity=0.8, draw opacity=1, text opacity=1, draw=white!80.00000!black},
            legend style={nodes={scale=0.8, transform shape}},
            legend style={draw=none},
            legend style={/tikz/every even column/.append style={column sep=0.5cm}}
            ]
            \addlegendimage{draw=color0, fill=color0, opacity=0.3, mark=*, only marks,mark size=2pt}
            \addlegendentry{Inactive devices ($\mathcal{K}\setminus \mathcal{K}_a$)};
            \addlegendimage{semithick, color1,mark=triangle*, only marks,mark size=2pt}
            \addlegendentry{Active devices ($\mathcal{K}_a$)};
            \addlegendimage{semithick, color2,mark=square*, only marks,mark size=2pt}
            \addlegendentry{Access points};
            
            \end{axis}
        \end{tikzpicture}
    \end{subfigure}\vspace{0.2cm}
    \begin{subfigure}[t]{0.45\linewidth}
        \input{img/co-located-map.tex}
        \centering
        \caption{Co-located case $N=1$.}
    \end{subfigure}\hfill%
    \begin{subfigure}[t]{0.45\linewidth}
		\input{img/cell-free-map.tex}
		\centering
        \caption{Cell-free case $N=32$.}\label{fig:initial-access-area-cell-free}
    \end{subfigure}
    \caption{Example of a simulation scenario with an area of $500 \times 500$~\si{\square\meter} for the case of $32$ total base station antennas for a co-located (a) and cell-free (b) deployment.}\label{fig:initial-access-area}
\end{figure}

\section{Motivation -- Recurrence in \gls{csi}}
\begin{figure}[hbtp]
    \centering
    \begin{subfigure}{0.8\linewidth}
        \centering
        \includegraphics[width=\linewidth]{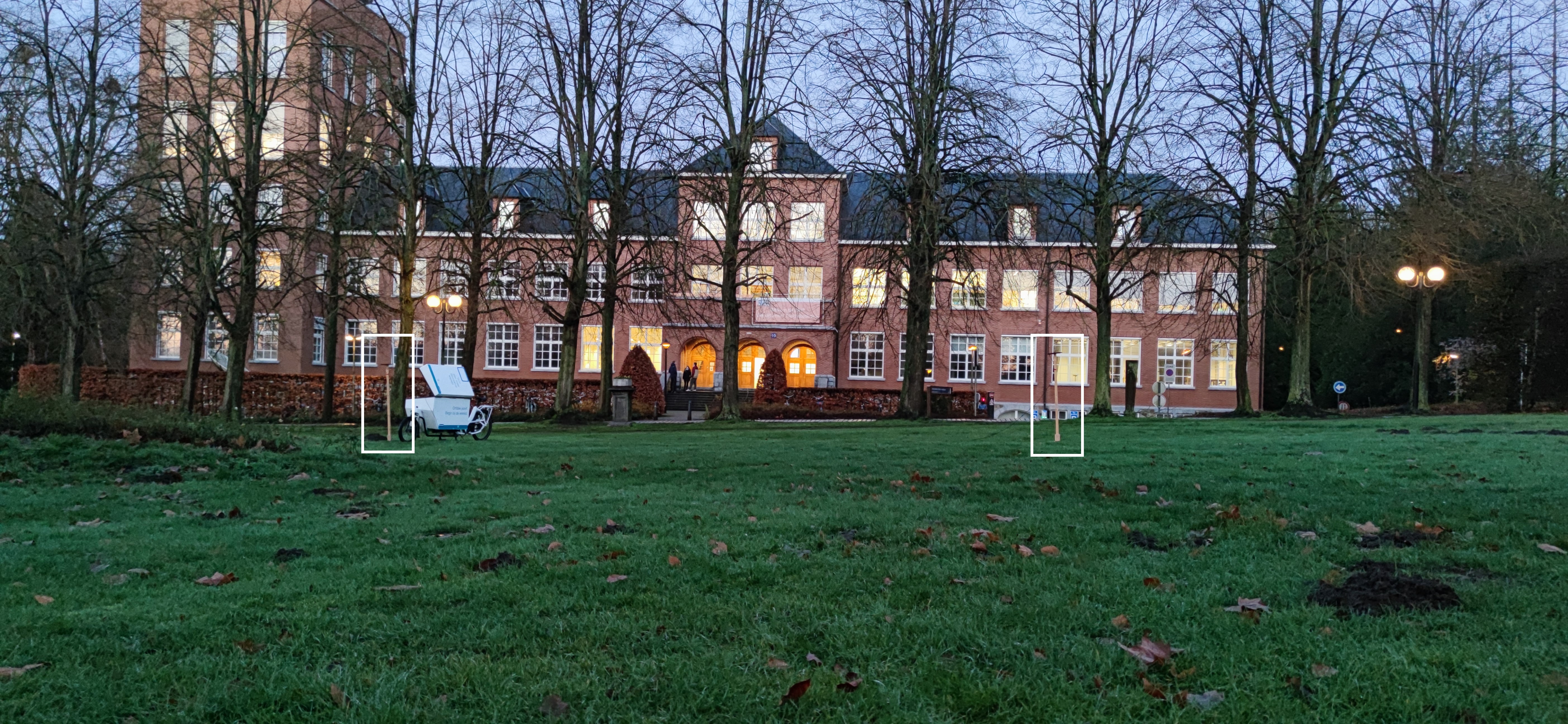}
    \caption{\Gls{ula} at the balcony with Node~1 positioned right and Node~2 left.}\label{fig:init-access-bike-and-building}
    \end{subfigure}

    \begin{subfigure}{0.8\linewidth}
        \centering
        \includegraphics[width=\linewidth]{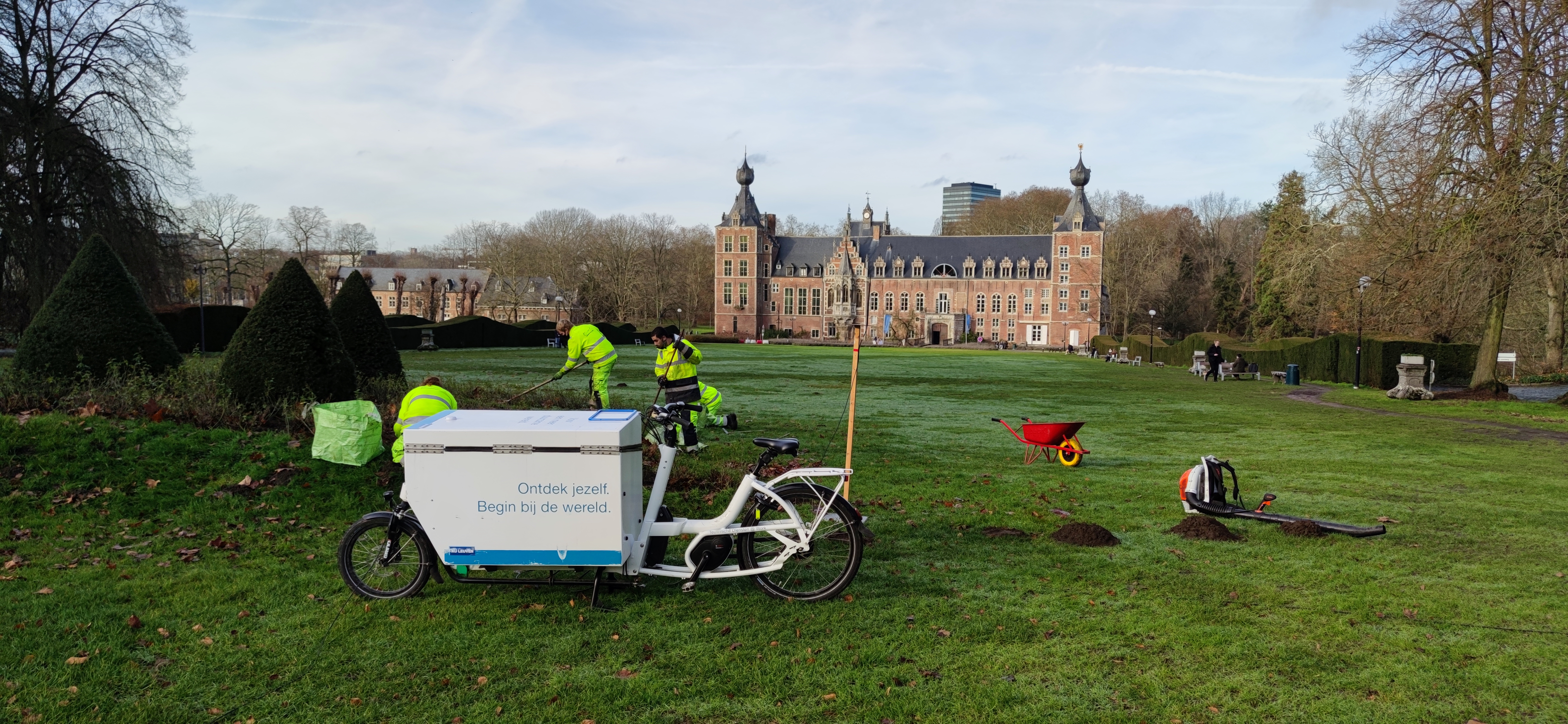}
    \caption{Movement in the proximity of Node~2 during the measurements.}\label{fig:init-access-bike-and-people}
    \end{subfigure}
    \caption{Measurement setup to study long-term channel behavior.}\label{fig:init-access-measurement-setup}
\end{figure}

\Gls{iot} -- and more specifically \gls{lpwan} -- technologies are often put in the field with a deploy-and-forget strategy, where the devices remain immobile afterwards. 
As such, we can expect that the channel conditions are less time-variant than assumed in theoretical models often assumed in literature~\cite{bjornsonBook,marzetta2016fundamentals}.  To investigate the long-term behavior of the channel, a \gls{ula}, described in~\cite{callebaut2021experimental}, is used to estimate the channel of two static \gls{iot} nodes, shown in Fig.~\ref{fig:init-access-measurement-setup}, over a time period of more than \num{8} hours. 
This measurement campaign represents a typical \gls{iot} scenario where the base station or gateway is located at an elevated height and the \gls{iot} devices are fixed in place. 
As shown in Fig.~\ref{fig:init-access-bike-and-people}, during the experiment, movement in the proximity of the nodes is present. Furthermore, the \gls{los} was sometimes blocked due to cars passing by, as would be the case in real deployments.

The long-term behavior is measured by taking the channel correlation (Eq.~\ref{eq:sub-mamimo-corr-coeff}) of the first channel estimate and the other channel $N$ measurements, i.e., $\delta_{1,j}\ \text{for}\ j \in \{1,\dotsc,N\}$.

\begin{equation}\label{eq:sub-mamimo-corr-coeff}
    \delta_{i,j} = 
    \frac
    {\abs{
    \hermconj{\normalized{\vectr{h}}_{i}} \cdot \normalized{\vectr{h}}_{j}
    }}
    {\norm{\normalized{\vectr{h}}_{i}} \norm{\normalized{\vectr{h}}_{j}}}
\end{equation}

\begin{figure}[hbtp]
    \centering
    \begin{subfigure}[t]{0.8\linewidth}
        \input{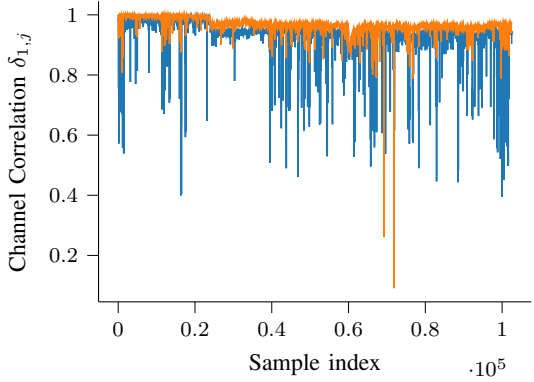}
        \caption{Channel correlation with respect to the first channel instance. Although the correlation sporadically drops, the majority of the channel instances are highly correlated, as shown in Fig.~\ref{fig:init-access-corr-kul-first-ecdf}.}\label{fig:init-access-corr-full-day}
    \end{subfigure}
    
    \begin{subfigure}[t]{0.8\linewidth}
		\input{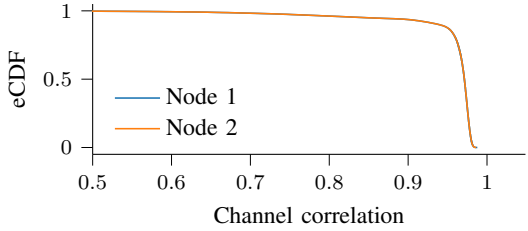}
        \caption{The \gls{ecdf} of the channel correlations. It indicates that 91\% for Node~1 and 95\% for Node~2 of the channel correlations are above \num{0.9}.}\label{fig:init-access-corr-kul-first-ecdf}
    \end{subfigure}
    \caption{Channel correlation over a full day (9h24-17h48) with over \num{10000} channel instances.}
\end{figure}

The observed channel correlations are depicted in Fig.~\ref{fig:init-access-corr-full-day}. The \gls{ecdf} of these correlation coefficients is shown in Fig.~\ref{fig:init-access-corr-kul-first-ecdf}.
Fig.~\ref{fig:init-access-corr-full-day} illustrates that most of the time the correlation coefficient is close to \num{1}, indicating that the channel is highly correlated with the first estimate and thus can be considered static. It also shows that, while in some occasions the correlation drops, the channel quickly becomes again highly correlated with the first channel instance. More than 90\% of the measured channels\footnote{More specifically, 91\% and 95\% for Node~1 and Node~2, respectively.} have a correlation coefficient higher than \num{0.9} over a window of more than \num{8} hours. This demonstrates the potential of re-using channel estimates in \gls{iot} contexts.\footnote{We use here the term ``re-using'' to indicate that we no longer operate in a block fading model with independent channel realizations. Typically, these blocks are considered in the order of \SI{50}{\milli\second}.} 

\section{system Model}
Two deployment strategies, i.e., co-located and cell-free massive MIMO, are considered, as shown in Fig.~\ref{fig:initial-access-area}.  The former follows the conventional massive MIMO where all antennas are located in one array, often spaced by half a wavelength. 
In cell-free systems, \glspl{ap} are geographically distributed over the area equipped with one or more antennas. In our study, we assume single-antenna \glspl{ap} in the cell-free case. In the remainder of the manuscript, we will use the term \gls{ap} to denote the base station (in the central case).

There are $K$ devices, each trying to access the network with a certain activity probability $\epsilon_a$. To do so, each active device $k\in\mathcal{K}_a$ sends a unique, non-orthogonal preamble of length $\tau_{p}$, known to the network. 
The pilot symbol of the preamble sent by device $k$ at pilot symbol $t$ is denoted by $s_{k,t}$. The vector of received symbols at the $M$ antennas of the \glspl{ap} at time $t$ is
\begin{align}
    \vectr{y}_t=\sum_{k=0}^{K-1}\vectr{g}_{k} s_{k,t} \gamma_k + \vectr{w}_t,
\end{align}
where $\vectr{g}_k \inC{M\times 1}$ is the known prior \gls{csi} of device $k$, $\gamma_k$ is an unknown complex scalar and $\vectr{w}_t\inC{M\times 1}$ is \gls{awgn}, distributed as $\cn{\vectr{0}}{\sigma^2 \matr{I}}$. 
The unknown complex scalar, $\gamma_k = \sqrt{\rho_k} a_k e^{j\phi_k}$, contains the transmit power $\rho_k$, device activity $a_k \in \left\{0,1\right\}$ and a potential phase offset $\phi_k$. The device-specific phase offset can account for a \gls{cfo}, where the \gls{cfo} is considered constant over the preamble duration. By assuming that all \(M\) antennas are perfectly synchronized, this offset is only dependent on the device. In case the device is inactive, $\gamma_k$ will be zero. 
Defining the matrices of channel vectors and pilot symbols at time $t$, 
$$\matr{G}=\left[\vectr{g}_{0}, \cdots ,\vectr{g}_{K-1}\right] \inC{M\times K},$$%
$$\matr{D}_t=\diag\left(s_{0,t},\cdots,s_{K-1,t}\right)\inC{K\times K},$$
the received signal at the $M$ base station antennas at time $t$ can be rewritten as
\begin{align}
\vectr{y}_t&=\matr{G} \matr{D}_t \vectr{\gamma} + \vectr{w}\\
\end{align}
The matrix $\matr{G} \matr{D}_t \inC{M\times K}$ is known at the \glspl{ap} as it consists of the known \gls{csi}, $\matr{G}$, and the transmitted preamble symbols from the $K$ devices, $\matr{D}_t$.
By stacking the $\tau_p$ consecutive $\vectr{y}_t \in \tilde{\vect{y}}$, the received signal becomes,
\begin{align}
	\tilde{\vect{y}}&=\begin{pmatrix}
	\mat{G}\mat{D}_{0}\\
	...\\
	\mat{G}\mat{D}_{\tau_p-1}
	\end{pmatrix}\vect{\gamma}+\tilde{\vect{w}}\\\nonumber
	&=\begin{pmatrix}
	\mat{G} & &\\
	& \ddots &\\
	& & \mat{G}
	\end{pmatrix}\begin{pmatrix}
	\mat{D}_{0}\\\nonumber
	...\\
	\mat{D}_{\tau_p-1}
	\end{pmatrix}\vect{\gamma}+\tilde{\vect{w}}\\\nonumber
	&=\left(\mat{I}_{\tau_p} \otimes \mat{G}\right)\mat{D}\vect{\gamma}+\tilde{\vect{w}}\\\nonumber
	&=\mat{\Gamma}\vect{\gamma}+\tilde{\vect{w}},
\end{align}
where $\mat{D} = \begin{pmatrix}
	\mat{D}_{0}\\\nonumber
	...\\
	\mat{D}_{\tau_p-1}
	\end{pmatrix}$ and $\mat{\Gamma} = \left(\mat{I}_{\tau_p} \otimes \mat{G}\right)\mat{D} \in \mathbb{C}^{M \tau_p \times K}$.

\section{Device Activity Detection}
A novel algorithmic solution is proposed to find the active devices. 
After determining the active devices, conventional detection techniques, e.g., \gls{zf}, can be used to separate the devices in the uplink.

As soon as $\tau_p \geq K/M$ and for well-conditioned channels and pilot sequences, the matrix $\mat{\Gamma}$ will be of full rank $K$ and $\vect{\gamma}$ can be estimated by using a left pseudo inverse,
\begin{align}
\hat{\vect{\gamma}}&=\left(\mat{\Gamma}^H\mat{\Gamma}\right)^{-1}\mat{\Gamma}^H \tilde{\vect{y}}\\ \nonumber
&= \vect{\gamma} + \left(\mat{\Gamma}^H\mat{\Gamma}\right)^{-1}\mat{\Gamma}^H\tilde{\vect{w}}.
\end{align}
This corresponds to the maximum likelihood estimate. 
Note that, when $M\tau_p \gg K$, $\mat{\Gamma}^H\mat{\Gamma}$ is expected to become a diagonal matrix because the device channels and unique preambles will become orthogonal~\cite{marzetta2016fundamentals}. This can lower the complexity of the inverse operation and allows for distributed processing.

A non-negative activity threshold $\gamma_{th,k}$ is applied for each device $k$. A device is considered active if $\abs{\hat{\gamma}_k} > \gamma_{th,k}$.
We define the real-valued threshold as,
\begin{equation}\label{eq:initial-access-threshold}
    \gamma_{th, k} = v \sqrt{\text{SNR}_k}^{-1},
\end{equation}
where $v$ is chosen to have a desired probability of false alarm and miss detection performance, as discussed in Section~\ref{sec:sims}. The \gls{snr} of device $k$, $\text{SNR}_k$, is $\rho_k \norm{\vectr{g}_k}^2 \norm{\vectr{s}_k}^2 / \sigma^2$, with $\sigma^2$ the noise power.

\section{Simulation Results}\label{sec:sims}
The performance of the proposed scheme is explored for a co-located and a cell-free system.  An example of a simulation scenario, with a co-located or cell-free \gls{ap} deployment, is shown in Fig.~\ref{fig:initial-access-area}. 
The number of \glspl{ap} is denoted by $N$. The default simulation configurations are summarized in Table~\ref{tab:initial-access-setup}. 

\begin{table}[htbp]
    \centering
    \caption{Simulation parameter set.}
    \label{tab:initial-access-setup}
    \centering
    \footnotesize
      \begin{threeparttable}
    \begin{tabular}{@{}lll@{}}
    \toprule
    Parameter &  Symbol & Default value\\ \midrule
    
    Number of devices & $K$ & \num{1000}\\
    Number of \glspl{ap} & $N$ & -\\
    Number of total \gls{ap} antennas & $M$ & -\\
    Area size & $A$ & $500 \times 500$ \si{\square\meter}\\
    Device activity probability & $\epsilon_a$ & \num{0.01}\\
    Transmit power of the device & $\rho$ & \SI{1}{\milli\watt}\\
    Bandwidth & $B$ & \SI{125}{\kilo\hertz}\\
    Thermal noise &$\sigma^2$ & \SI{-122.88}{\dBm} \\
    Path loss & $\beta$ & urban~model\cite{7377400}\\
    Carrier Frequency & $f_c$& \SI{868}{\mega\hertz}\\
    Pilot sequence & $s_k$ & $\cn{0}{1}$\\
    Pilot length &$\tau_p$  & \num{40} symbols\\
    \Gls{iot} device height & $h_k$ & \SIrange{1}{4}{\meter}\\
    \Gls{ap} height & $h_{\text{BS}}$ & \SI{29}{\meter}\\
    Phase offset & $\phi_k$ &$\sim\mathcal{U}_{[0, 2\pi]}$\\
    Number of simulations & $N_\text{sim}$ & \num{1000}\\
    \bottomrule
    \end{tabular}
    \end{threeparttable}
    \end{table}

We consider an area of $500 \times 500$\,\si{\square\meter} with \num{1000} devices. The positions of the devices are randomly generated for each simulation. The locations are kept constant among the different scenarios in order to have a fair comparison. The locations of the access points are generated in order to have, on average, a uniform distribution of the \gls{ap} positions, as shown in Fig.~\ref{fig:initial-access-area-cell-free}. For each simulation, a grid of size $\left \lceil\sqrt{N}\, \right \rceil$~by~$\left \lceil \sqrt{N}\, \right \rceil$ is generated of which $N$ random positions are selected for the \glspl{ap}. In case there is only one \gls{ap}, the \gls{ap} is located in the center.

The device activity profile is generated randomly and independently for each device with a probability $\epsilon_a=0.01$, meaning that on average \num{10} devices are active simultaneously. Or equivalently, the devices have an average duty cycle of 1\%, which can be considered high~\cite{s21030913} for the investigated applications and hence represents a worst-case scenario. A higher number of simultaneously active devices is simulated by increasing the activity probability $\epsilon_a$. 
All devices use the same transmit power of \SI{1}{\milli\watt}. We assume the same bandwidth as used in \gls{lorawan}, i.e., \SI{125}{\kilo\hertz}, and have adopted the thermal noise floor for a \SI{125}{\kilo\hertz} signal, i.e., $\sigma^2 = \SI{-122.88}{\dBm}$, at the receiver. The channel between \gls{ap} $n$ and device $k$ is modeled as $\vect{g}_{k,n} =  \sqrt{\beta_{k,n}} h$, with $\beta_{k,n}$ the path loss and $h$ the small-scale fading \gls{iid} $\cn{0}{1}$. The large-scale \acrlong{pl}  follows the reported model in~\cite{7377400} for an urban environment operating at \SI{868}{\mega\hertz}, i.e., $\beta_{k,n}= 128.95 + 23.2\log_{10}(d_{k,n}) + \chi$~\si{\deci\bel} with $d_{k,n}$ the distance between device $k$ and \gls{ap} $n$ and $\chi$ the shadow fading$\sim\mathcal{N}\left(0,7.8\right)$. For the co-located case,  $\beta_{k,n}$ becomes $\beta_{k}$.

In the proposed scheme, we consider that the channel $\matr{G}$ is static in time.
The pilot sequence is randomly generated from a complex Gaussian distribution $s_k \cn{0}{1}$, and is assumed to be known by all \glspl{ap}. Each device uses a pilot sequence of \num{40} symbols, respecting the requirement of $\tau_p$ being greater than or equal to $K/M$. A random phase offset $\phi_k \sim\mathcal{U}_{[0, 2\pi]}$ is generated to simulate a \acrlong{cfo} (considered time-invariant over the simulation period).


\subsubsection{Trading-off the probability of false alarm and miss detection}
The \gls{roc} is studied based on the probability of miss detection and false alarm.
False alarm happens when a device is considered active, while it was actually not transmitting. In contrast, a miss detection occurs if a device was active but is undetected.
As in~\cite{9154288}, the probability of miss detection is defined as the average ratio of undetected devices to the number of active devices
\begin{equation}
    \prob_{md} = 1 - \expt{
            \dfrac{
                \abs{\mathcal{K}_a \cap \hat{\mathcal{K}}_a }
            }{
                \abs{\mathcal{K}_a}
            }
        },
\end{equation}
where $\mathcal{K}_a$ is the set of active devices and $\hat{\mathcal{K}}_a = \{k | \hat{a}_k=1, \forall \in [1,K]\}$ denotes the estimated set of active devices.
Note that on average $\abs{\mathcal{K}_a}=K_a$.
Similarly, the probability of false alarm is the ratio of inactive devices considered active to the number of inactive devices and is given by
\begin{equation}
    \prob_{fa} = \expt{
            \dfrac{
                \abs{\hat{\mathcal{K}}_a \setminus  \mathcal{K}_a}
            }{
                K - \abs{\mathcal{K}_a}
            }
    }.
\end{equation}

\begin{figure}[h]
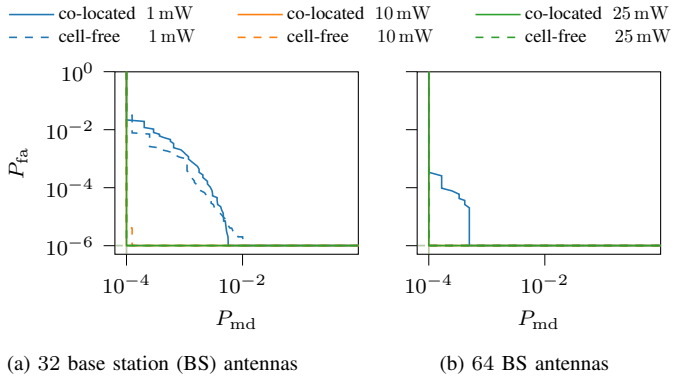

    \centering
    \begin{subfigure}{\linewidth}
        \centering

        \begin{tikzpicture} 
            \definecolor{color0}{rgb}{0.12156862745098,0.466666666666667,0.705882352941177}
            \definecolor{color1}{rgb}{1,0.498039215686275,0.0549019607843137}
            \definecolor{color2}{rgb}{0.172549019607843,0.627450980392157,0.172549019607843}
            \begin{axis}[%
            hide axis,
            xmin=10,
            xmax=50,
            ymin=0,
            ymax=0.4,
            legend style={draw=white!15!black,legend cell align=left},
            legend columns=3,
            legend cell align={left},
            legend style={fill opacity=0.8, draw opacity=1, text opacity=1, draw=white!80.00000!black},
            legend style={nodes={scale=0.7, transform shape}},
            legend style={draw=none},
            legend style={/tikz/every even column/.append style={column sep=0.5cm}}
            ]
            \addlegendimage{semithick, color0}
            \addlegendentry{co-located\hspace{0.2cm}\SI{1}{\milli\watt}};

            \addlegendimage{semithick, color1}
            \addlegendentry{co-located\hspace{0.2cm}\SI{10}{\milli\watt}};
            \addlegendimage{semithick, color2}
            \addlegendentry{co-located\hspace{0.2cm}\SI{25}{\milli\watt}};

            \addlegendimage{semithick, color0, dashed}
            \addlegendentry{cell-free\hspace{0.535cm}\SI{1}{\milli\watt}};
            \addlegendimage{semithick, color1, dashed}
            \addlegendentry{cell-free\hspace{0.535cm}\SI{10}{\milli\watt}};
            \addlegendimage{semithick, color2, dashed}
            \addlegendentry{cell-free\hspace{0.535cm}\SI{25}{\milli\watt}};
            \end{axis}
        \end{tikzpicture}
    \end{subfigure}
    \begin{subfigure}[t]{0.45\linewidth}
        \centering
        \input{img/powers-M32.tex}%
        \centering%
        \caption{\num{32} \gls{bs} antennas}
    \end{subfigure}\hfill%
    \begin{subfigure}[t]{0.45\linewidth}
        \centering
        \input{img/powers-M64.tex}%
        \centering%
        \caption{\num{64} \gls{bs} antennas}
    \end{subfigure}%
    \caption{\Gls{roc} for \num{32} and \num{64} base station antennas for different transmit power. By using only \num{64} antennas in a cell-free system, the \gls{iot} devices can lower their transmit power to \SI{1}{\milli\watt} or \SI{0}{\dBm}.}\label{fig:long-term-csi-powers}
\end{figure}

\begin{figure}[h]
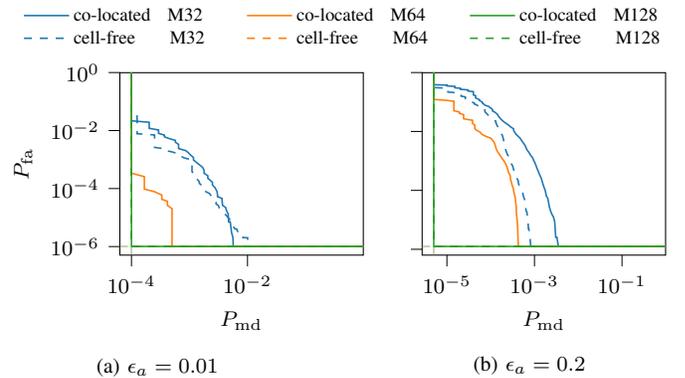

    \centering
    \begin{subfigure}{\linewidth}
    \centering
        \begin{tikzpicture} 
            \definecolor{color0}{rgb}{0.12156862745098,0.466666666666667,0.705882352941177}
            \definecolor{color1}{rgb}{1,0.498039215686275,0.0549019607843137}
            \definecolor{color2}{rgb}{0.172549019607843,0.627450980392157,0.172549019607843}
            \begin{axis}[%
            hide axis,
            xmin=10,
            xmax=50,
            ymin=0,
            ymax=0.4,
            legend style={draw=white!15!black,legend cell align=left},
            legend columns=3,
            legend cell align={left},
            legend style={fill opacity=0.8, draw opacity=1, text opacity=1, draw=white!80.00000!black},
            legend style={nodes={scale=0.7, transform shape}},
            legend style={draw=none},
            legend style={/tikz/every even column/.append style={column sep=0.5cm}}
            ]
            \addlegendimage{semithick, color0}
            \addlegendentry{co-located\hspace{0.2cm} M32};

            \addlegendimage{semithick, color1}
            \addlegendentry{co-located\hspace{0.2cm} M64};
            \addlegendimage{semithick, color2}
            \addlegendentry{co-located\hspace{0.2cm} M128};

            \addlegendimage{semithick, color0, dashed}
            \addlegendentry{cell-free\hspace{0.535cm} M32};
            \addlegendimage{semithick, color1, dashed}
            \addlegendentry{cell-free\hspace{0.535cm} M64};
            \addlegendimage{semithick, color2, dashed}
            \addlegendentry{cell-free\hspace{0.535cm} M128};
            \end{axis}
        \end{tikzpicture}
    \end{subfigure}\newline%
    \begin{subfigure}[t]{0.45\linewidth}
        \centering
        \input{img/prob-E1.tex}%
        \vspace{-0.4cm}
        \caption{$\epsilon_a=0.01$} %
    \end{subfigure}\hfill%
    \begin{subfigure}[t]{0.45\linewidth}
        \centering
        \input{img/prob-E20.tex}%
        \caption{$\epsilon_a=0.2$} %
    \end{subfigure}
    \caption{The co-located system experiences more performance degradation than cell-free systems when increasing the activity probability. Even with 20\% of the devices being active, the systems using \num{64} (cell-free) and \num{128} (both cell-free and co-located) \gls{ap} antennas do not detect any errors. Notice that the x-axis is scaled differently as the minimal observable probability of miss detection is lower for the $\epsilon_a=0.2$ case due to a higher number of active devices, which is on average $200$ opposed to $10$ for $\epsilon_a=0.01$.}\label{fig:long-term-csi-active-prob}
\end{figure}

The results presented in Fig.~\ref{fig:long-term-csi-powers}, \ref{fig:long-term-csi-area-size}, and \ref{fig:long-term-csi-active-prob}, are scaled to the lowest non-zero probability measurable, which depends on the number of nodes, active nodes $\abs{\mathcal{K}_a}$ and number of simulations $N_\text{sim}$. 
The minimum obtained probability of miss detection is observed if only one device of all active devices is undetected, i.e., $ 1/ \left(\sum^{N_\text{sim}}_{i=1} \abs{\mathcal{K}_{a,i}}\right) \approx 1 / (K_a * N_\text{sim})$.
Equivalently, the lowest false alarm is perceived when only one inactive device is considered active, i.e., $1 / \left(K N_\text{sim} - \sum^{N_\text{sim}}_{i=1} \abs{\mathcal{K}_{a,i}}\right) \approx  1 / ((K-K_a) * N_\text{sim})$.
For the default case with \num{1000} simulations, this becomes on average $10^{-6}$ and $10^{-4}$ for the probability of false alarm ($\prob_\mathrm{fa}$) and miss detection ($\prob_\mathrm{md}$), respectively.
A trade-off can be made between the two probabilities by varying $v$ in~(\ref{eq:initial-access-threshold}). A lower $v$ yields a lower activity threshold, resulting in more devices considered active. This in turn lowers the probability of miss detection, while increasing the probability of generating a false alarm. In the simulations, the parameter $v$ is swept across the range $[10^{-2}, 10^{5}]$. In the remainder of the manuscript, the optimal threshold $v_{\text{opt}}$ is used to denote the choice of $v$ yielding the lowest probability of false alarm and miss detection. Due to the limits imposed by the minimum probabilities observable, there exists a region where no errors of false alarm and miss detection, were observed. 

In the following, the impact of varying the device transmit power, area size and activity level on the device detection performance is studied.


\paragraph{Effect of the transmit power}  

The impact of the devices' transmit power (\SI{1}{\milli\watt}, \SI{10}{\milli\watt} and \SI{25}{\milli\watt}) when using \num{32} and \num{64} antennas is shown in Fig.~\ref{fig:long-term-csi-powers}. 
In the case of \num{128}~antennas, no false alarm or miss detection has been observed when using the optimal threshold $v_{\text{opt}}$ for all three transmit powers. 
For both \SI{10}{\milli\watt} and \SI{25}{\milli\watt} transmit power the cell-free and co-located case, even with only \num{32} antennas, perform optimal. 
In case of using \SI{1}{\milli\watt} transmit power, \num{32} antennas show not to be sufficient. As of \num{64}~antennas only in the co-located system, miss detection and false alarm are observed, but the probabilities are reduced by a factor of \num{100}.
This demonstrates that the transmit power of \gls{iot} devices could be greatly reduced by moving to a cell-free scenario, while still requiring only a limited number of \gls{ap} antennas.

\paragraph{Effect of the area size}
\begin{figure}[h]
    \centering
    \begin{subfigure}{\linewidth}
        \centering

        \begin{tikzpicture} 
            \definecolor{color0}{rgb}{0.12156862745098,0.466666666666667,0.705882352941177}
            \definecolor{color1}{rgb}{1,0.498039215686275,0.0549019607843137}
            \definecolor{color2}{rgb}{0.172549019607843,0.627450980392157,0.172549019607843}
            \begin{axis}[%
            hide axis,
            xmin=10,
            xmax=50,
            ymin=0,
            ymax=0.4,
            legend style={draw=white!15!black,legend cell align=left},
            legend columns=3,
            legend cell align={left},
            legend style={fill opacity=0.8, draw opacity=1, text opacity=1, draw=white!80.00000!black},
            legend style={nodes={scale=0.7, transform shape}},
            legend style={draw=none},
            legend style={/tikz/every even column/.append style={column sep=0.5cm}}
            ]
            \addlegendimage{semithick, color0}
            \addlegendentry{co-located\hspace{0.2cm} M32};

            \addlegendimage{semithick, color1}
            \addlegendentry{co-located\hspace{0.2cm} M64};
            \addlegendimage{semithick, color2}
            \addlegendentry{co-located\hspace{0.2cm} M128};

            \addlegendimage{semithick, color0, dashed}
            \addlegendentry{cell-free\hspace{0.535cm} M32};
            \addlegendimage{semithick, color1, dashed}
            \addlegendentry{cell-free\hspace{0.535cm} M64};
            \addlegendimage{semithick, color2, dashed}
            \addlegendentry{cell-free\hspace{0.535cm} M128};
            \end{axis}
        \end{tikzpicture}
    \end{subfigure}

    \begin{subfigure}[t]{0.45\linewidth}
        \centering
        \input{img/area-A500.tex}%
        \centering%
        \caption{$500\times500$\,\si{\meter\squared} area}
    \end{subfigure}\hfill%
    \begin{subfigure}[t]{0.45\linewidth}
        \centering%
        \input{img/area-A1000.tex}%
        \centering%
        \caption{$1000\times1000$\,\si{\meter\squared} area}
    \end{subfigure}
    
    \caption{\Gls{roc} for a $500\times500$\,\si{\meter\squared} and $1000\times1000$\,\si{\meter\squared} area. The cell-free systems experience a lower performance degradation due to the higher \gls{pl} of the $1000\times1000$\,\si{\meter\squared} area.}\label{fig:long-term-csi-area-size}
\end{figure}
Fig.~\ref{fig:long-term-csi-area-size} illustrates the impact of the area size on the performance of the initial access scheme for a $500\times500$ and a $1000\times1000$\,\si{\squared\meter} area size.
The first obvious effect is the degradation of the \gls{snr} due to a higher path loss of the larger area, yielding a decrease in performance for all scenarios. The cell-free system is less affected by the increase in area size, as also observed in~\cite{9154288}. 

\paragraph{Effect of the activity level} Fig.~\ref{fig:long-term-csi-active-prob} shows the impact of an increase in activity probability from \num{0.01} to \num{0.2}. The latter implies that, on average, \num{200} devices simultaneously access the network. 
Similar to previous observations, the cell-free systems outperform the co-located networks. Despite this, even with 20\% of the devices being active on average, the systems using \num{64} (cell-free) \num{128} (both cell-free and co-located) \gls{ap} antennas, do not detect any errors.

\section{Conclusions}
A grant-free algorithm is proposed using prior \gls{csi} for initial access in massive MIMO networks. It serves as a baseline for the potential of re-using \acrlong{csi} to support massive and low-power \gls{iot}. 
The initial scheme has a closed form solution, without the need for iterations, is scalable to a large number of devices, and is robust to \gls{cfo}. 

The performance of the proposed algorithm is studied for a conventional co-located and cell-free system. A trade-off between the probability of false alarm and miss detection can be made. The simulation results demonstrate that with only \SI{1}{\milli\watt} of transmit power, \num{64} base station antennas, deployed in a cell-free setup, will suffice to support a large area ($1000\times1000$ \si{\squared\meter}) and a high number of access requests (up to $200$) and still having a probability of false alarm and miss detection below $10^{-6}$ and $10^{-4}$, respectively. Furthermore, the results also indicate that we could greatly reduce the transmit power of \gls{iot} devices by moving to a cell-free scenario, while requiring only a limited number of total \gls{ap} antennas.

In contrast to what is observed in~\cite{senel2018grant}, with our algorithm, the performance of a cell-free topology performs better than a co-located case. 
This is because we are able to exploit the full \acrlong{csi} and thus also the increase in channel diversity due to the geographical distribution of the \glspl{ap}.

As an extension, the model and algorithm could be adapted to include partial \gls{csi}, as opposed to perfectly knowing the full \acrlong{csi}. In addition, currently, single-antenna \glspl{ap} are considered for the cell-free case. The simulations can be extended by investigating the optimal number of \glspl{ap} for a fixed number of total receive antennas. 

{\footnotesize%
\bibliographystyle{IEEEtranN}%
\bibliography{./bibliography/bib}%
}

\end{document}

%% file: abbr.tex
\newacronym{iot}{IoT}{Internet of Things}
\newacronym{lorawan}{LoRaWAN}{long-range wide-area network}
\newacronym{lora}{LoRa}{long range}
\newacronym{css}{CSS}{chirp spread spectrum}
\newacronym{lpwan}{LPWAN}{low-power wide-area network}
\newacronym{fsk}{FSK}{frequency shift keying}
\newacronym{adr}{ADR}{adaptive data rate}
\newacronym{mac}{MAC}{medium access control}
\newacronym{abp}{ABP}{activation-by-personalisation}
\newacronym{otaa}{OTAA}{over-the-air activation}
\newacronym{api}{API}{application programming interface}
\newacronym{aes}{AES}{advanced encryption standard}
\newacronym{ldo}{LDO}{low-dropout regulator}
\newacronym{ide}{IDE}{integrated development environment}
\newacronym{sf}{SF}{spreading factor}
\newacronym{rf}{RF}{radio frequency}
\newacronym[plural=PAs,firstplural=power amplifiers (PAs)]{pa}{PA}{power amplifier}
\newacronym{bw}{BW}{bandwidth}
\newacronym{mcu}{MCU}{microcontroller unit}
\newacronym{spi}{SPI}{serial peripheral interface}
\newacronym{los}{LoS}{line-of-sight}
\newacronym{kpi}{KPI}{key performance indicator}
\newacronym{gprs}{GPRS}{general packet radio services}
\newacronym{egprs}{EGPRS}{enhanced data rates for GSM evolution}
\newacronym{nbiot}{NB-IoT}{Narrowband IoT}
\newacronym{lte}{LTE}{long term evolution}
\newacronym{lpwa}{LPWA}{low-power wide-area}
\newacronym{m2m}{M2M}{machine-to-machine}
\newacronym{cbm}{CBM}{condition based maintenance}
\newacronym{ai}{AI}{artificial intelligence}
\newacronym{edrx}{eDRX}{extended discontinuous reception mode}
\newacronym{drx}{DRX}{discontinuous reception mode}
\newacronym{psm}{PSM}{power saving mode}
\newacronym{gps}{GPS}{global positioning system}
\newacronym{ptw}{PTW}{paging time window}
\newacronym{tau}{TAU}{tracking area update}
\newacronym{mtc}{MTC}{machine-type communication}
\newacronym{ue}{UE}{user equipment}
\newacronym{cdrx}{cDRX}{connected mode DRX}
\newacronym{3gpp}{3GPP}{3rd generation partnership project}
\newacronym{gsm}{GSM}{global system for mobile communications}
\newacronym{prbs}{PRBs}{physical resource blocks}
\newacronym{mcl}{MCL}{maximum coupling loss}
\newacronym{ce}{CE}{coverage enhancement}
\newacronym{rsrp}{RSRP}{reference signals received power}
\newacronym{rtc}{RTC}{real time clock}
\newacronym{ism}{ISM}{industrial, scientific and medical}
\newacronym{mmtc}{MMTC}{massive machine type communication}
\newacronym{scfdma}{SC-FDMA}{single-carrier frequency division multiple access}
\newacronym{udp}{UDP}{user datagram protocol}
\newacronym{snr}{SNR}{signal-to-noise ratio}
\newacronym{sinr}{SINR}{signal-to-interference-plus-noise ratio}
\newacronym{der}{DER}{data extraction rate}
\newacronym{dr}{DR}{data rate}
\newacronym{p2p}{P2P}{point-to-point}
\newacronym{rss}{RSS}{received signal strength}
\newacronym{erp}{ERP}{effective radiated power}
\newacronym{lbt}{LBT}{listen before talk}
\newacronym{afa}{AFA}{adaptive frequency agility}
\newacronym{pdr}{PDR}{packet delivery ratio}
\newacronym{pl}{PL}{path loss}
\newacronym{ml}{ML}{maximum-likelihood}
\newacronym{ols}{OLS}{ordinary least squares}
\newacronym{cdf}{CDF}{cumulative distribution function}
\newacronym{pdf}{PDF}{probability density function}
\newacronym{cobyla}{COBYLA}{constrained optimization by linear approximation}
\newacronym{per}{PER}{packet error rate}
\newacronym{mae}{MAE}{mean absolute error}
\newacronym{rmse}{RMSE}{root mean square error}
\newacronym{cr}{CR}{coding rate}
\newacronym{ofdm}{OFDM}{orthogonal frequency-division multiplexing}
\newacronym{ofdma}{OFDMA}{orthogonal frequency division multiple access}
\newacronym{qos}{QoS}{quality of service}
\newacronym{ra}{RA}{random access}
\newacronym{rar}{RAR}{random access response}
\newacronym{ttm}{TTM}{time to market}
\newacronym{cots}{COTS}{commercial off-the-shelf}
\newacronym{multi-rat}{multi-RAT}{multiple radio access technology}
\newacronym{lib}{LIB}{lithium-ion battery}
\newacronym{lipo}{LIPO}{lithium polymer}
\newacronym{msk}{MSK}{minimum-shift keying}
\newacronym{qam}{QAM}{quadrature amplitude modulation}
\newacronym{em}{EM}{electromagnetic}
\newacronym{papr}{PAPR}{peak-to-average power ratio}
\newacronym{pv}{PV}{photovoltaic}
\newacronym{dbpsk}{DBPSK}{differential binary phase shift keying}
\newacronym{bpsk}{BPSK}{binary phase shift keying}
\newacronym{qpsk}{QPSK}{quadrature phase shift keying}
\newacronym{pmu}{PMU}{power management unit}
\newacronym{bod}{BOD}{brown-out detection}
\newacronym{ram}{RAM}{random-access memory}
\newacronym{io}{IO}{input/output}
\newacronym{fram}{FRAM}{ferroelectric random access memory}
\newacronym{fft}{FFT}{fast fourier transform}
\newacronym{tpms}{TPMS}{tire-pressure monitoring system}
\newacronym{phy}{PHY}{physical}
\newacronym{pc}{PC}{lithium poly carbon}
\newacronym{edlc}{EDLC}{electric double-layer capacitor}
\newacronym{cmos}{CMOS}{complementary metal-oxide-semiconductor}
\newacronym{soc}{SoC}{state of charge}
\newacronym{SoC}{SOC}{system on chip}
\newacronym{imu}{IMU}{inertial measurement unit}
\newacronym{agps}{A-GPS}{assisted global positioning system}
\newacronym{i2c}{I2C}{inter-integrated circuit}
\newacronym{adc}{ADC}{analog-to-digital converter}
\newacronym{led}{LED}{light emitting diode}
\newacronym{gpio}{GPIO}{general-purpose input/output}
\newacronym{pcb}{PCB}{printed circuit board}
\newacronym{prs}{PRS}{peripheral reflex system}
\newacronym{sha}{SHA}{secure hash algorithm}
\newacronym{ecc}{ECC}{elliptic-curve cryptography}
\newacronym[plural=UVs,firstplural=Unmanned Vehicles]{uv}{UV}{unmanned vehicle}
\newacronym{voc}{VOC}{voltatile organic compound}
\newacronym{iaq}{IAQ}{indoor air quality}
\newacronym{gnss}{GNSS}{global navigation satellite system}
\newacronym{ttff}{TTFF}{time to first fix}
\newacronym{pll}{PLL}{phase-locked loop}
\newacronym{crc}{CRC}{cycle redundancy check}
\newacronym{dma}{DMA}{direct memory access}
\newacronym{cpu}{CPU}{central processing unit}
\newacronym{dmac}{DMAC}{direct memory access controller}
\newacronym{dram}{DRAM}{dynamic random-access memory}
\newacronym{mppt}{MPPT}{maximum power point tracking}
\newacronym{sram}{SRAM}{static random-access memory}
\newacronym{uart}{UART}{universal asynchronous receiver/transmitter}
\newacronym{awgn}{AWGN}{additive white gaussian noise}
\newacronym{at}{AT}{ATtention}
\newacronym{pse}{PSE}{power sourcing equipment}
\newacronym{pd}{PD}{powered device}
\newacronym{usrp}{USRP}{universal software radio peripheral}
\newacronym{rpi}{RPi}{Raspberry Pi}
\newacronym{poe}{PoE}{power over ethernet}
\newacronym{ptp}{PTP}{precision time protocol}
\newacronym{vlc}{VLC}{visible light communication}
\newacronym{vlp}{VLP}{visible light positioning}
\newacronym{tsn}{TSN}{time-sensitive networking}
\newacronym{sdr}{SDR}{software-defined radio}
\newacronym{pps}{PPS}{pulse per second}
\newacronym{gpclk}{GPCLK}{general purpose clock}
\newacronym{ros}{ROS}{robot operating system}
\newacronym{tof}{ToF}{time-of-flight}
\newacronym{ntp}{NTP}{network time protocol}
\newacronym{uhd}{UHD}{USRP hardware driver}
\newacronym{os}{OS}{operating system}
\newacronym{hat}{HAT}{hardware attached on top}
\newacronym{gpu}{GPU}{graphical processing unit}
\newacronym{if}{IF}{intermediate frequency}
\newacronym{fpga}{FPGA}{field-programmable gate array}
\newacronym{rfic}{RFIC}{radio-frequency integrated circuit}
\newacronym{ssd}{SSD}{solid state drive}
\newacronym{mimo}{MIMO}{multiple input multiple output}
\newacronym{mumimo}{MU-MIMO}{multiuser MIMO}
\newacronym[plural=BSs,firstplural=base stations (BSs)]{bs}{BS}{base station}
\newacronym{ib}{IB}{in-band}
\newacronym{oob}{OOB}{out-of-band}
\newacronym{rts}{RTS}{ray-tracing simulator}
\newacronym{nlos}{NLoS}{non line-of-sight}
\newacronym{evm}{EVM}{error vector magnitude}
\newacronym{mr}{MR}{maximum ratio}
\newacronym{zf}{ZF}{zero forcing}
\newacronym{aclr}{ACLR}{adjacent channel leakage power ratio}
\newacronym{bo}{BO}{back-off}
\newacronym{psd}{PSD}{power spectral density}
\newacronym{im}{IM}{intermodulation}
\newacronym{arp}{ARP}{antenna reference point}
\newacronym{ag}{AG}{array gain}
\newacronym{ula}{ULA}{uniform linear array}
\newacronym{rrc}{RRC}{radio resource connection}
\newacronym{ecdf}{eCDF}{empirical cumulative distribution function}
\newacronym{iid}{i.i.d.}{independently and identically distributed}
\newacronym{hpbm}{HPBM}{half-power beam width}
\newacronym{mf}{MF}{matched filtering}
\newacronym{mrt}{MRT}{maximum ratio transmission}
\newacronym{rx}{RX}{receiver}
\newacronym{trp}{TRP}{transmission reception point}
\newacronym{tx}{TX}{transmitter}
\newacronym{dl}{DL}{downlink}
\newacronym{ul}{UL}{uplink}
\newacronym{rzf}{RZF}{regularized zero forcing}
\newacronym{enb}{eNB}{evolved Node B}
\newacronym{rat}{RAT}{radio access technology}
\newacronym{csi}{CSI}{channel state information}
\newacronym{dsp}{DSP}{digital signal processing}
\newacronym{tdma}{TDMA}{time-devision multiple-access}
\newacronym{ota}{OTA}{over-the-air}
\newacronym{ura}{URA}{uniform rectangular array}
\newacronym{ap}{AP}{access point}
\newacronym{dsss}{DSSS}{direct-sequence spread spectrum}
\newacronym{tdd}{TDD}{time-division duplexing}
\newacronym{fdd}{FDD}{frequency-division duplexing}
\newacronym{omp}{OMP}{orthogonal matching pursuit}
\newacronym{sucre}{SUCRe}{strongest-user collision resolution}
\newacronym{mse}{MSE}{mean-square error}
\newacronym{cfo}{CFO}{carrier frequency offset}
\newacronym{roc}{ROC}{receiver operating characteristic}
\newacronym{dop}{DOP}{dilution of precision}
\newacronym{nnls}{NNLS}{non-negative least squares}
\newacronym{mmv-amp}{MMV-AMP}{multiple measurement vector - approximated massage passing}

%% file: img/co-located-map.tex
\begin{tikzpicture}
            \definecolor{color0}{rgb}{0.12156862745098,0.466666666666667,0.705882352941177}
            \definecolor{color1}{rgb}{1,0.498039215686275,0.0549019607843137}
            \definecolor{color2}{rgb}{0.172549019607843,0.627450980392157,0.172549019607843}
    
    \begin{axis}[
    tick align=outside,
    tick pos=left,
    x grid style={white!69.0196078431373!black},
    xmin=-25, xmax=525,
    xtick style={color=black},
    y grid style={white!69.0196078431373!black},
    ymin=-25, ymax=525,
    ytick style={color=black},
    width=0.8\linewidth,
    height=0.8\linewidth,
    scale only axis,    legend cell align={left},
    legend style={fill opacity=0.8, draw opacity=1, text opacity=1, draw=white!80.00000!black},
    legend style={nodes={scale=0.7, transform shape}},
    legend style={draw=none},
    ]
    \addplot [semithick, draw=color2, fill=color2,mark=square*, only marks, mark size=1.5pt]
    table [row sep=\\]{%
    250.0 250.0\\
    };\addlegendentry{Access points}
    \addplot [draw=color0, fill=color0, mark=*, only marks, opacity=0.1]
    table [row sep=\\]{%
    x  y
    230.334371782779 229.658163126284\\
    276.127936574795 274.950963950937\\
    418.785765593361 91.4076003318218\\
    258.137246123375 325.142748422535\\
    60.2921859595403 162.567241332486\\
    432.288409908053 468.532137609445\\
    426.154484774662 84.5353992703653\\
    141.458806039821 270.996676471303\\
    475.490445659766 480.37533276801\\
    450.609940419705 159.116680410221\\
    340.754419878455 64.9384630931073\\
    454.730875844874 245.662760779903\\
    120.863634457726 318.044675595344\\
    144.070822865166 85.4331419728776\\
    482.821964054575 470.439207781853\\
    448.723650725267 313.213095364414\\
    22.1517884046958 153.997613739249\\
    147.572708981291 499.287759776016\\
    91.2043498883721 245.441326338527\\
    69.4877447784283 372.527792662632\\
    109.97642157878 143.906883077292\\
    401.224674326341 150.818798909608\\
    133.085434091911 329.563619512531\\
    77.2465903032352 91.2991570463059\\
    465.51471980842 2.19849795679516\\
    469.829361844642 280.233358583632\\
    485.513425631963 472.38214464312\\
    363.360872161183 378.241079231682\\
    365.732252866519 256.014182831287\\
    36.6340285248462 388.58164680901\\
    114.133629649203 160.983825626773\\
    441.568848608022 369.891661808094\\
    250.497765738214 325.886296917112\\
    323.906194793287 191.719765394961\\
    62.3692662608432 90.8081522500862\\
    58.1295071136498 360.964196157812\\
    365.363394588264 411.371190840284\\
    59.7350754720869 472.77821008412\\
    395.56284498658 114.840932742899\\
    372.507254113088 64.3904610858189\\
    106.933548146224 486.32994628173\\
    150.095401650491 409.51219512758\\
    154.465050483524 197.807947907705\\
    73.860108602356 112.299533221101\\
    142.686118973439 122.274026902067\\
    248.224786951449 136.153778158126\\
    20.3412919061315 15.6273700194509\\
    380.638551228761 491.696396293417\\
    395.827407058873 205.374450145835\\
    277.74724786363 151.297272900449\\
    217.852613436892 483.485443016232\\
    182.009700950796 461.235444275993\\
    366.403616540931 277.823369108692\\
    326.657696317318 495.934599074243\\
    233.08243245269 181.512768997491\\
    47.0669941620931 81.0622609302099\\
    434.752631910077 295.10364152989\\
    219.670823611993 163.983624489277\\
    370.378224422928 57.7419890708681\\
    324.706127114456 251.031542270877\\
    63.8637150998588 371.939058566304\\
    270.568531653515 238.479629184072\\
    63.3281427579969 395.588559233737\\
    23.4120762432324 434.21535329231\\
    347.741857016181 50.4990533321926\\
    106.279142140497 286.177990975054\\
    9.31266935384106 487.713349001768\\
    182.788898632879 402.141374808115\\
    363.558064621095 247.38707239626\\
    53.2773852054858 263.338147938154\\
    74.3328605706963 33.0654903495625\\
    464.963903330382 478.607439907609\\
    27.2653990087082 332.158857532779\\
    137.167658905608 124.02236550571\\
    225.746479446037 18.5335457439736\\
    50.1535336122066 263.774550115159\\
    256.35972728377 365.935883681716\\
    379.978447792752 303.933399936026\\
    463.292469636297 58.0531946073926\\
    285.62104691036 30.4231221696135\\
    411.972291245246 439.886321326354\\
    219.831310031148 95.2235792876249\\
    437.445724586301 123.24050292189\\
    89.3680619376633 441.754516166309\\
    241.257048066147 468.659886172886\\
    241.353735056472 417.062478869489\\
    255.765100913287 139.313000385597\\
    473.187285916074 367.791409131051\\
    18.3175462753328 497.652826374767\\
    47.0218867806086 459.029242397728\\
    443.527550809082 412.675271962832\\
    375.021684767399 463.956050243738\\
    406.336317326663 376.026747406053\\
    97.346835616713 278.828374453608\\
    307.512280216105 441.962802360933\\
    200.257770378306 422.267070401602\\
    33.9369806925682 357.866351230504\\
    102.145941581931 129.299001495282\\
    83.5192166802122 404.409672546479\\
    228.479477095805 182.368982855973\\
    73.1612461862952 381.388664408206\\
    391.886607187981 91.0948514162951\\
    295.815822616956 355.430358089875\\
    481.738014961433 274.688176190992\\
    60.680238417199 417.562789680489\\
    256.117841636017 273.25107901337\\
    398.935191213885 26.5283742638234\\
    462.144333287043 144.39249629801\\
    55.940264772001 355.701251528573\\
    405.369792524079 454.427174024094\\
    0.272446963420658 37.9140219265426\\
    100.699650012338 484.941112512727\\
    58.6064198063263 112.554173322132\\
    368.342891019078 86.1950825307584\\
    308.235717178149 75.7886616322613\\
    466.962407372968 447.208497263062\\
    162.104593774472 299.245685082418\\
    59.5623733908272 330.8898267659\\
    391.721398108864 346.977934259957\\
    402.2736554693 349.623967540541\\
    164.708445172172 195.380141259444\\
    320.446329068191 370.596837101341\\
    480.051192156681 358.947194333562\\
    431.029516442483 469.49064019035\\
    76.1828950066023 147.855708840668\\
    154.05950687034 270.536158014355\\
    282.795000693726 99.2910232227179\\
    392.78250585021 393.274482036307\\
    72.654934848618 332.094921135013\\
    368.602853387657 476.431210517195\\
    265.337642728876 392.47879082254\\
    476.14139952637 212.001260542258\\
    463.88708944035 215.985012678597\\
    89.6443651010327 370.049716318686\\
    410.182658474704 2.55328166167473\\
    343.609748295597 63.1613574242176\\
    155.028517300474 208.475845493589\\
    177.185242420911 116.276919633716\\
    476.634584695593 328.892389911033\\
    305.803525391578 28.9640357842683\\
    0.378621008883029 23.5503879710811\\
    10.8347816839194 382.914196082519\\
    408.929933511548 140.096957537079\\
    478.822957576732 350.17580317261\\
    471.797288110117 49.941499794323\\
    191.1511830362 373.407404180066\\
    3.53967007879097 421.726460245155\\
    41.7965934093249 284.399586223175\\
    376.432347753157 322.663123743897\\
    266.920543992353 72.5594450962321\\
    426.462985611756 459.017178423617\\
    180.326925699742 3.7537870983459\\
    258.653278550675 96.410081128033\\
    455.654797829634 127.85768445487\\
    444.648673536594 495.11842648113\\
    365.008905961045 261.213518851178\\
    197.174587185071 328.543598330698\\
    265.910288347206 88.747665269823\\
    386.245590704275 467.698572900258\\
    444.971501987037 170.553653075884\\
    361.008572728094 363.051519155093\\
    488.262584005742 286.08841505185\\
    433.596243026601 77.0372959600055\\
    128.147424390062 383.929105342975\\
    385.151457102256 142.869542056847\\
    92.2346414395868 378.461630939791\\
    291.755978645085 418.442893769882\\
    11.4932270502738 25.075887004655\\
    262.442224600726 271.859072221259\\
    204.283100576423 267.882461856167\\
    242.928299595269 63.4455641231591\\
    235.968145554414 35.5703638000995\\
    319.186008594579 27.216910060602\\
    408.342189129246 461.791098427187\\
    316.02777102199 74.0397894795248\\
    131.776993115145 321.087927908794\\
    204.901254167196 471.550567332733\\
    229.247154897556 75.4168137427073\\
    348.153956712368 56.5004259641323\\
    126.090825249525 168.64882175829\\
    35.2565725791379 33.0462609757969\\
    429.339192724354 26.7725636651915\\
    450.915272577523 459.233235539839\\
    276.032849551205 205.484045695385\\
    426.253979148184 204.130354879611\\
    293.62629302835 417.382546552569\\
    210.403016285395 45.0414634324631\\
    220.329832965304 497.618531428907\\
    19.6763561439786 204.050052812284\\
    249.880294124428 57.8247800520291\\
    116.82392624238 489.308590918627\\
    495.395312937344 154.433869250428\\
    133.323969476472 402.061612846143\\
    493.845097264141 248.704307643412\\
    68.5383837847698 34.9431799004293\\
    443.938028354247 330.964710066555\\
    46.4662402602514 283.110233870002\\
    76.4173892381232 2.89718339219053\\
    216.783299517268 149.664129519472\\
    154.024349734127 389.744392131155\\
    345.158254225426 156.006968864174\\
    292.995611694811 373.167288020707\\
    276.806349437205 351.934586183662\\
    12.6502536713929 491.632703495278\\
    477.758807598436 285.55376175013\\
    45.2736054780359 293.52444517893\\
    280.448513459086 141.480658284369\\
    146.407313238409 159.182051578613\\
    393.004352211542 0.812956167566325\\
    104.135302692941 460.774815928997\\
    345.725295812487 347.035242899397\\
    156.903612328356 418.176941593216\\
    354.603489720881 39.7527327405764\\
    203.416755158029 100.159716276543\\
    181.77439806228 221.271115292828\\
    101.694317696418 112.765250986341\\
    228.058794309985 90.9484701884102\\
    173.069357795223 429.959188158895\\
    426.730715633196 277.290614236043\\
    487.398884394392 213.217205458547\\
    438.309439681715 429.596819707505\\
    497.098603398645 58.0442967137685\\
    151.415952680487 382.866238620498\\
    86.6882826902952 205.959691244491\\
    422.291731763618 220.445522123469\\
    8.04807968375781 291.572805631621\\
    268.991907088113 76.190463998728\\
    273.852057359903 10.6583589030569\\
    418.928602980972 21.7477977205611\\
    480.779332713886 292.567929501634\\
    240.448852709863 282.355400764788\\
    44.7932804831399 222.721224141543\\
    254.174339442859 435.8292397566\\
    386.999745703782 397.302593494521\\
    470.42469200646 238.277528434523\\
    279.544098467624 300.298044834172\\
    485.532289342294 322.099003206017\\
    468.403727595025 123.766555684276\\
    289.09857445754 345.187613503981\\
    62.6242168369271 359.091150370236\\
    426.431262038621 431.611633312447\\
    217.494283874309 238.634323757128\\
    104.007245839379 191.797034969695\\
    203.888867880197 206.491785737417\\
    220.391942302841 162.614988063084\\
    324.588373820955 499.154822018112\\
    445.185319642341 150.010794259308\\
    189.15909839878 76.2823523178276\\
    176.002085766477 102.602758068251\\
    127.008387164458 396.70676626766\\
    458.483147522455 158.983750619556\\
    99.3027468911721 447.547655492994\\
    335.915485674062 151.806039668577\\
    148.376396557313 351.469146313992\\
    193.727120299708 203.246868806576\\
    1.90827414429962 244.196066647327\\
    271.471127976884 136.907857268178\\
    14.4488186167075 422.305681945899\\
    45.3586781575613 235.907015174207\\
    90.0913195936377 39.8732043751722\\
    163.533549005172 15.865220102561\\
    164.146402777546 250.747501102713\\
    181.603699484528 51.473368096855\\
    284.856923367219 168.659661635049\\
    362.660873786492 272.905769037288\\
    411.499452765468 463.47519111877\\
    201.506378558326 471.372359742329\\
    298.519415444749 333.11222360294\\
    455.55389114742 211.456260178687\\
    401.116196434727 169.322775647102\\
    437.362990909153 466.432100879868\\
    40.9722587348114 485.694390324735\\
    123.370094007119 117.26349816936\\
    87.8256403634233 460.081860845924\\
    325.623860065416 240.871731314639\\
    14.5446308899832 335.81776726045\\
    318.175304296699 397.028402903851\\
    123.803565875205 187.099151783001\\
    254.763792270727 298.485788995258\\
    363.097336680844 141.039790668592\\
    392.558650168972 270.436530670064\\
    249.716116439744 448.165812411041\\
    459.632173336458 307.65532419955\\
    126.63457382109 247.114916604083\\
    1.13663122321006 35.7359479046061\\
    407.163350924516 334.229074361364\\
    192.574782271769 245.9699999071\\
    498.309149873147 245.738233521224\\
    273.415485901719 499.540727213266\\
    144.757874277679 226.862343874721\\
    10.4228711799573 309.442128894789\\
    54.662197348671 427.042267044867\\
    415.220141440814 297.292594239848\\
    100.961291272675 152.064464513022\\
    336.378405014999 78.942414228637\\
    336.962201470663 417.493472007369\\
    27.86825992967 179.537143408907\\
    22.7969237668469 24.9681362673652\\
    58.7883264922893 19.2823542765266\\
    318.718264972533 133.321734577247\\
    346.618129949346 390.0188783495\\
    86.3426753014914 222.735917572245\\
    352.958150448585 75.9179485152801\\
    152.320063032784 3.09894020821583\\
    71.4122623182641 335.896920526867\\
    358.930023927155 410.158906009803\\
    323.507300749364 435.154118579919\\
    428.15030768338 19.158689255965\\
    45.1738608665622 90.3927346011665\\
    464.268147156035 185.788589702156\\
    256.620728698177 158.84770230259\\
    416.335997227721 136.062879907615\\
    240.941144661804 263.280149119155\\
    384.47024000595 60.1170921631858\\
    120.904209863949 250.652281121505\\
    86.559063606799 460.142119926363\\
    423.311711741138 86.9238576254883\\
    2.60376640679311 153.832969555133\\
    136.993236188047 309.386051500792\\
    403.387082698616 42.77135339968\\
    349.922741102947 90.4230996917377\\
    121.191071114406 485.152279650117\\
    60.5094913857941 54.2446543759556\\
    481.256848335314 138.538286317653\\
    328.690095057579 216.940602924311\\
    472.505453126243 146.758760405178\\
    95.8513962772983 466.605190297031\\
    292.943106748094 387.661672254777\\
    293.32894460275 454.387961016807\\
    36.1899134978801 223.845541479839\\
    430.449526163641 270.746043353223\\
    336.932910342125 447.845791526155\\
    380.069132074195 226.610911046093\\
    466.99726915268 208.192757080887\\
    111.056529052537 115.251176828181\\
    62.9973232463695 11.555875037406\\
    27.2891774977888 183.22723954364\\
    401.070075362191 286.898866195256\\
    359.080881225841 118.286188068486\\
    99.0796537727492 484.685247334006\\
    417.393900099844 411.454747298179\\
    204.486020026141 227.939073958929\\
    304.241169990389 184.73209129827\\
    60.96203838895 249.703642206093\\
    331.468175571561 28.2581872425065\\
    499.9456775868 134.900251112982\\
    264.229310081275 377.047807834306\\
    113.552549434178 314.149892122409\\
    421.373709315519 255.518697708452\\
    378.728785535828 365.353973877083\\
    351.306768284033 304.830141321346\\
    277.196813900792 253.865620305548\\
    465.470691915917 322.037173491556\\
    95.5203921720639 403.524024709267\\
    77.4458051799253 288.981433966368\\
    464.102793312517 316.638134430004\\
    74.193854807392 241.207465173113\\
    418.992840220393 193.357708326874\\
    184.019689556145 302.315071927867\\
    228.19605195384 9.57071572396462\\
    472.396844435735 227.044949068707\\
    328.758488197195 492.079058681402\\
    345.883692438664 257.768500238076\\
    402.747572545138 95.7926955244308\\
    495.782286718262 480.837931848851\\
    148.542500031413 301.437973401857\\
    258.209290784321 100.268576024666\\
    313.895723848676 387.135359016927\\
    63.0631922365245 42.8387321612512\\
    397.367533301081 283.912005558171\\
    287.74196870496 71.5740298890724\\
    200.049129025739 326.670855540801\\
    112.518432171703 429.731780729899\\
    107.520292754536 410.496163746502\\
    57.2443190065537 42.4496926601553\\
    350.274663301403 85.7795600163211\\
    200.360767367044 231.453075794505\\
    447.084709961588 74.7306247916925\\
    191.999001395543 458.11093183166\\
    248.359575447356 163.230970289465\\
    20.5042684212365 363.246818664278\\
    248.718745732034 136.565195667762\\
    313.729514257329 40.0879158707818\\
    97.4454799227119 317.08966541498\\
    83.9966193726703 499.544916870037\\
    98.4275271947832 432.565300025795\\
    428.726886343129 389.937160884782\\
    431.356370675505 349.831657198785\\
    151.700857921428 432.894259591963\\
    254.840537180012 399.950982579525\\
    470.534250096678 101.036847163621\\
    427.412745623024 175.280666340019\\
    127.179683588962 468.782315103552\\
    214.915150283908 328.924202645531\\
    356.735143677256 166.525896790244\\
    467.076401120849 271.729234738076\\
    151.537127677041 206.627642787837\\
    112.789686108012 399.917612079894\\
    410.788797401929 277.289473140683\\
    197.495808370415 196.714564707061\\
    279.28438257969 417.203146188078\\
    343.681070774048 189.208829442948\\
    97.0940878577286 230.096117433547\\
    335.211698176753 223.002451931118\\
    410.336889433903 141.598713285334\\
    476.10128903218 290.772765779076\\
    84.6255198746449 398.156011592846\\
    374.17781473239 332.067713197384\\
    199.455207017176 186.458114601936\\
    63.5797941745555 304.224676296139\\
    469.816350827095 447.394964520419\\
    368.061612103459 239.584082932599\\
    433.280486053445 472.777180151736\\
    207.016568895874 440.369208010751\\
    293.666682651537 365.218344747278\\
    364.160833590462 375.325833648317\\
    262.192835754898 163.447987871005\\
    380.975980096996 448.442391813999\\
    391.632339513928 176.952155703322\\
    13.2327337233236 4.08086901184429\\
    283.297038639038 123.170897944401\\
    205.065319025253 133.945984926157\\
    91.4843192779597 0.821207278592695\\
    225.481399429388 99.7380742423415\\
    255.66217653068 479.024481053067\\
    325.946839568393 237.985879261108\\
    72.4077124643764 484.462111078597\\
    384.535039812078 164.43668664712\\
    412.373244266754 393.231486607253\\
    272.0057175752 258.678216759218\\
    166.137084777068 176.729281607413\\
    306.089276904843 295.276392422497\\
    169.063332673594 112.706863591958\\
    336.606946143349 125.076866646639\\
    477.037354750096 407.434644805357\\
    313.281540222178 362.463738976112\\
    436.279313342579 298.136202524883\\
    321.198273597564 498.961508384998\\
    499.600138575464 104.006002570057\\
    240.561038691837 295.16742639337\\
    76.494837853682 139.873368862426\\
    494.131536333844 89.9924487856304\\
    86.0090606833642 1.86293588441855\\
    357.651546439624 279.837829092479\\
    420.407602525836 363.918809844994\\
    484.934583940738 358.01105077694\\
    249.239494478512 62.6126423691801\\
    199.430197851778 402.529289141575\\
    189.275503876651 373.216351306942\\
    104.93177026085 490.865133056387\\
    20.7746612378166 461.618587889119\\
    355.447080861458 5.43089172694622\\
    338.012366398729 395.761421961836\\
    97.506051142187 166.622980758624\\
    191.604431670783 190.795620108131\\
    396.497989714963 314.072702730295\\
    71.086349239718 14.6607330585214\\
    431.853977555933 364.333662228565\\
    460.677187090644 483.743575432131\\
    34.6172395449261 358.804847663158\\
    423.759025468613 126.801127279156\\
    247.685911494976 22.6773071707206\\
    372.782019247487 241.638631797094\\
    197.247860123563 329.552474855122\\
    428.096141734913 289.011954625744\\
    18.4236852725387 311.315457431876\\
    49.3162649001074 43.0320490867326\\
    137.306459177901 252.015776673926\\
    33.522042540286 25.7815590043991\\
    382.542012115407 68.350900716766\\
    213.371003615289 20.670627398315\\
    182.245919698751 29.2769892741873\\
    482.9407331817 227.722370699333\\
    167.39030326544 292.096066702021\\
    284.271335754327 215.682354555882\\
    450.857976601719 445.813388247692\\
    152.361252684243 364.448157413336\\
    394.312188006776 4.96533561165491\\
    272.310585848313 418.780994371999\\
    254.146522900541 435.122150505596\\
    92.6567132943707 177.281255842727\\
    91.0794262769919 178.733046859993\\
    257.453112856832 33.1038238956787\\
    62.7548056569631 251.39122824241\\
    91.5039219399045 169.635961682734\\
    234.460527695316 186.378533990824\\
    260.888795844826 352.751142808158\\
    206.716728598542 49.8216844323334\\
    9.42963636693922 218.676843062069\\
    168.479690546406 18.3934415533596\\
    99.4749825863598 24.4028881411173\\
    469.03722362976 269.114343342971\\
    313.526761855946 384.672080582989\\
    345.628546122784 93.149816829786\\
    152.520236501768 325.790056069292\\
    132.529985675338 91.8294834216131\\
    410.873815587418 320.631115717882\\
    168.039906408767 458.448255578457\\
    302.993744278228 310.461150035517\\
    170.775686693078 23.8748050676239\\
    304.736415167667 297.131738811075\\
    122.55563776323 440.516180669231\\
    129.883588969131 290.321786358056\\
    385.510022925188 351.124892603358\\
    129.288109876252 123.22075059756\\
    70.6220907156006 459.684421544316\\
    223.616819717394 20.8867397219618\\
    290.358160425711 466.254099100954\\
    339.290132866772 440.419294316937\\
    454.457432359202 446.294644664717\\
    278.901356981207 357.990900375717\\
    460.805971364807 439.476949622451\\
    86.9406888683164 438.889341131468\\
    322.321274712481 16.9113674187008\\
    154.293453360174 420.93622736633\\
    441.581482265659 38.3062189051349\\
    443.579132600868 129.230121056561\\
    12.5189668207118 221.358957870478\\
    415.274552842847 248.670525176238\\
    475.338234826965 300.202117176352\\
    387.077335016815 382.96001682567\\
    195.580382229661 173.837057336554\\
    282.758893030291 167.315845396248\\
    115.910210441828 32.0672435416255\\
    265.322178731347 218.635858883949\\
    376.627439228509 41.6385403197799\\
    209.666122229001 270.886075634049\\
    394.270965469065 155.941999099846\\
    158.671294199798 9.31110345686331\\
    124.87053511956 44.7396853714928\\
    311.216468288144 73.4924412301661\\
    249.417128990457 417.018827779818\\
    483.657143572591 373.198798562666\\
    457.901084868955 397.245953312734\\
    428.715258940648 257.253779782766\\
    207.489641466073 248.151190015492\\
    156.327424451711 85.8143474705125\\
    77.9129477618715 396.68114372921\\
    45.4070282614247 56.3319656536861\\
    89.081422725878 413.394731573902\\
    302.883576869806 425.965080688404\\
    30.9020137899043 440.338254794455\\
    255.826300443825 33.3790113153648\\
    74.1423901547801 410.244258685477\\
    302.834949442203 438.052353918077\\
    476.42870097441 430.074418867574\\
    416.567571235842 264.984662333554\\
    366.100659540948 353.569111202935\\
    261.655255678217 327.261330579807\\
    394.85651334715 441.453527135054\\
    308.790042942292 277.164373263785\\
    323.166436766421 444.753614818009\\
    84.7427750738616 378.837991327281\\
    403.144950206051 136.638040161647\\
    111.428974918088 245.654231837422\\
    55.2954646360655 472.498759728045\\
    112.596956400873 277.99181625717\\
    253.936680269003 364.429439633529\\
    116.977387414029 254.115454471435\\
    116.702991998787 461.218224266087\\
    364.557890691295 38.8196676610633\\
    146.613442478245 99.7390496883767\\
    114.155283226648 369.953313758736\\
    118.976885794845 273.593582104944\\
    170.273421907271 449.364031972215\\
    281.020420335399 104.779529912708\\
    354.51567268505 225.753149313624\\
    206.272895814716 260.480721224011\\
    299.065578719556 88.4572134639826\\
    429.147294242418 276.122117633375\\
    301.237318005074 22.4951383749689\\
    442.516990042866 236.090317726719\\
    160.549247704769 355.896376121005\\
    10.6618167733585 257.436733434655\\
    388.144299331249 297.947308569313\\
    371.622431046919 150.172523661899\\
    465.319231932176 190.065918386169\\
    164.878623373057 398.040156295367\\
    460.376780965122 247.111793937615\\
    50.3308590054002 38.2366309994845\\
    94.9186392119644 159.878133910178\\
    66.3398005831694 490.058575739707\\
    49.1576749559245 357.290285160844\\
    17.9240894557608 37.4540530988633\\
    219.327092214832 163.789477291217\\
    377.185085110356 329.418873145238\\
    494.528033504284 198.473720565433\\
    115.086971707281 359.926507146286\\
    409.462545815263 14.5222145761144\\
    6.79376421320793 329.001655369942\\
    88.8513885334264 403.574198413886\\
    234.958149003508 417.912005803844\\
    51.5690942511067 286.668178617932\\
    407.748890553167 362.692035241822\\
    239.72004990793 116.194113296253\\
    204.011161378654 247.233687577587\\
    135.258484529991 69.945640210788\\
    410.029644948416 0.722548523617139\\
    250.165335464365 231.813156064149\\
    141.894343848419 282.732720494229\\
    411.580137877465 135.518261514067\\
    259.037135761369 372.83232710426\\
    418.236819062714 454.889976171566\\
    53.1187401482863 28.2170231167049\\
    381.065563241943 363.215396288029\\
    449.87158772878 41.6443640288399\\
    223.636658211238 215.82772053373\\
    102.637678922373 103.050673330147\\
    5.75036192054829 491.113776131868\\
    238.640860493246 149.975983435738\\
    204.639160567514 140.130481202486\\
    7.17334217723853 196.681682762283\\
    70.9704084310049 28.4517881989729\\
    219.400607515389 260.727737973644\\
    142.950263551393 291.820912719496\\
    54.8127300989341 224.658358160791\\
    146.974202748304 338.068457233154\\
    466.95766445648 367.617143310442\\
    344.568928063753 1.13959710041006\\
    274.601065983703 294.848932905896\\
    243.530772726906 366.380406072125\\
    448.47582071203 280.217582587976\\
    424.695639985854 52.4450390468817\\
    357.62729734412 464.937963378679\\
    298.675673036866 179.27894844702\\
    270.822874052071 109.566449858281\\
    366.497539444872 3.04016464905338\\
    10.8316247381112 253.97835304387\\
    271.74856006147 247.998996158568\\
    125.221779933981 192.893835415468\\
    355.991829683548 435.435133817447\\
    425.883300936487 201.392485235782\\
    360.606422422793 176.825139086723\\
    88.36790652865 413.36606144983\\
    81.7493752883062 360.761550845745\\
    223.359417241621 130.745969528354\\
    203.076451737736 227.192695853279\\
    333.019348317558 246.165229649005\\
    311.062022889012 197.607579693486\\
    136.943185393171 354.604541524963\\
    132.07350843189 380.656998742825\\
    455.331503103858 407.446339969648\\
    29.7696899143189 91.2797604882228\\
    361.952103096624 171.25171242567\\
    208.961899723023 157.21871719068\\
    199.075363756919 374.882813576191\\
    150.375862584026 67.8917430458952\\
    268.421336888703 493.212035964632\\
    397.651223360278 294.459163818319\\
    402.042762344609 7.05752590612702\\
    387.412249320192 460.471590685213\\
    484.823469045488 256.265547613208\\
    113.184354847619 385.998365716677\\
    447.020054870679 384.803781029374\\
    287.666055048804 156.734067027295\\
    141.534129867629 275.652110176992\\
    244.435697048488 200.565860891158\\
    424.955960774324 212.400571345514\\
    196.547083817694 498.616107984368\\
    228.647032369692 496.032074879246\\
    252.539939023146 315.956403646606\\
    160.671592218136 29.9067900403389\\
    140.013625209693 252.257350368665\\
    188.817383505563 61.2977229589518\\
    258.613543903207 26.89213879668\\
    499.552446760012 182.254578329777\\
    80.4342838532666 81.4458469102726\\
    9.66437985150886 441.466569800765\\
    325.413685485616 470.699343837153\\
    278.552066257535 411.659297569882\\
    452.971419661761 315.185565458496\\
    244.023099313913 191.864303637413\\
    459.476961960265 45.0888289987454\\
    16.626029315719 224.977692295722\\
    387.631545205931 419.912450134257\\
    403.922580964848 113.587405488484\\
    463.795229272965 311.498857514017\\
    148.841732778572 426.680324830588\\
    388.859300878882 212.503290207789\\
    380.319002603798 149.4403668774\\
    344.195202750616 278.70021363808\\
    353.024184123784 128.315505327487\\
    336.24433525023 364.120181510889\\
    313.96757656994 130.426217773423\\
    262.248842793892 396.204693966226\\
    59.5177451697938 456.282692906734\\
    391.995990102316 50.0591763845718\\
    378.773377728712 268.065000997547\\
    234.875136329515 91.3286283654803\\
    458.822215613317 305.761533023149\\
    214.568526217936 226.760882162949\\
    76.4938711175617 413.919917740723\\
    268.643910545002 470.108802652589\\
    253.401658492631 411.747732095327\\
    395.209757618334 438.146119444567\\
    351.088211054831 62.9465896406791\\
    223.436042898553 83.3349090303822\\
    11.5355892478439 44.4029725184532\\
    319.040477738471 375.558580649971\\
    361.035753051443 248.75047315269\\
    389.928524822231 85.2780016174208\\
    451.980221166012 378.68250794405\\
    257.609425871833 243.859890886924\\
    324.556846899068 43.7624157195775\\
    269.84661459407 215.204146383572\\
    182.838322153892 361.364062005838\\
    132.555701854065 349.935401936275\\
    146.96681935343 383.481791814396\\
    10.2183039998891 213.444030080857\\
    364.599257878207 266.606204232984\\
    255.434236015291 133.735187574397\\
    260.245564694593 207.981610487125\\
    148.419836333394 215.019655196399\\
    143.706999456785 239.600658412893\\
    282.76685036663 378.067934212063\\
    231.38614850165 279.144209096244\\
    11.8094849843322 25.5956484806581\\
    378.604814044552 68.1594879629358\\
    302.6993185789 358.453802483438\\
    18.7077311083565 11.6346135640826\\
    332.391594263717 117.957483071052\\
    328.725205239196 278.78780284751\\
    75.4056183099671 21.5735046815225\\
    146.550544917977 336.79590009174\\
    360.233741189262 471.901386882779\\
    356.697544407565 422.2386606351\\
    156.649356644395 110.529046551063\\
    118.928587430975 484.377647862901\\
    64.9549020101737 182.055527131071\\
    258.588505039456 165.450636670678\\
    89.8864481781199 405.88972359045\\
    30.9825868812748 377.832284635691\\
    262.52726866832 156.655583951783\\
    348.51998857475 436.095316271804\\
    199.267662514411 248.642435025004\\
    308.3842769033 408.164586934088\\
    201.823513011845 182.451610366237\\
    4.20580537488707 324.90005361136\\
    365.071664733413 107.646022789712\\
    355.311478458005 488.823172261508\\
    248.501753964084 209.91559530341\\
    189.396349008439 34.3705919422284\\
    66.4187753914961 422.203177411943\\
    431.229706589471 14.2665438610435\\
    109.151349356028 228.534265498074\\
    481.383362857083 67.1394076841864\\
    1.74111499231372 334.067988173499\\
    296.439806895895 361.52008324006\\
    146.971839920332 410.925221555424\\
    179.278200539604 490.379372156643\\
    431.234257761026 99.1739720863137\\
    484.234023515501 366.124500596844\\
    25.7030584422015 112.129242048795\\
    89.6426361606763 331.973581311632\\
    433.552084105249 11.8897544999703\\
    395.509345611328 59.3531830411618\\
    422.278489755848 59.8268870894273\\
    420.174237409198 431.176754099119\\
    382.935230855501 302.913678088358\\
    133.259937654625 390.398426948387\\
    130.539144885867 328.242269703339\\
    65.6868473097534 297.66314963778\\
    406.942863516031 76.0357398098415\\
    121.821506875278 66.0178228994483\\
    137.789416425434 142.105605377501\\
    221.056058895422 77.3393152389458\\
    293.435110545292 178.309771594326\\
    231.032731870165 69.3924382937395\\
    361.765742503936 458.993586633006\\
    357.158561087919 230.430899041971\\
    341.977763892559 128.870777437342\\
    40.0425396468054 280.450115962617\\
    428.271402469684 162.47393548204\\
    167.327857325078 87.3412735647954\\
    111.347951824388 487.438226933504\\
    497.052942091855 136.529029332451\\
    61.5651378755996 315.436341908862\\
    445.69953244009 31.8890126016262\\
    289.792272739318 132.332570150122\\
    490.214653164036 216.895359875399\\
    40.5315225942229 87.5655908198234\\
    319.858872744572 272.615303962327\\
    377.050354402829 368.811077276193\\
    212.168566888836 422.691536334577\\
    431.62772178195 164.262071273693\\
    371.791056656899 43.8951428803275\\
    272.999439094579 417.629032528677\\
    399.44276198582 320.743494182871\\
    253.359519133383 342.151424769678\\
    103.571598841205 10.9349048349317\\
    496.056582564567 105.8467288613\\
    55.7409728386087 481.446377724931\\
    368.238585947091 79.3159831938782\\
    107.199400420626 150.291123753881\\
    304.234559343172 376.66247888607\\
    136.252981957482 95.8657880809406\\
    283.977550818831 4.74501586918113\\
    325.077620013174 409.74810121406\\
    372.903947365976 218.56725783618\\
    498.276072341265 101.61901100486\\
    396.083237451207 122.511711178612\\
    87.773558397534 327.721524814276\\
    86.8807673501469 278.544973344374\\
    63.4458653788014 425.54470296794\\
    40.8465914868966 222.051021743359\\
    278.234234903249 325.77367069565\\
    420.35345751118 223.230458122302\\
    79.1443774099709 417.50595072921\\
    403.919781933584 450.181471072113\\
    123.469777799902 291.451347634124\\
    398.648410248982 389.240949554592\\
    163.284545607035 205.957507204285\\
    95.801810903006 291.462181614896\\
    343.372514955166 180.230707985475\\
    199.221043251758 13.6727315816791\\
    75.6275919703936 496.810978277317\\
    246.004223992852 398.522641150814\\
    51.9519190265165 268.66066789364\\
    394.640541653542 182.176788688616\\
    105.522889830389 315.465428809941\\
    282.552050016749 394.251379786259\\
    383.271278143609 453.58622198514\\
    419.921486112785 241.628575589159\\
    407.959239451748 32.4060845833857\\
    4.70580820352806 364.703410509919\\
    495.501978885698 353.545629768281\\
    318.696069213317 381.45571678444\\
    326.553116305388 404.539929948942\\
    100.689503468991 439.014090594459\\
    463.103613637543 162.473153663017\\
    442.164015595153 386.143552596075\\
    276.66490072406 381.57414305058\\
    491.411611608058 132.940375498561\\
    422.502304844671 425.492533525794\\
    487.573208273725 274.980509871934\\
    451.417207770078 51.6373036683337\\
    395.080986881901 175.838284940426\\
    269.588193222371 295.798193066261\\
    459.030451007108 317.6860064187\\
    154.874425823668 285.864407587977\\
    9.70744857533973 396.789458470347\\
    357.304879012271 123.642993690618\\
    72.2798513267278 344.908286399603\\
    453.335335458946 94.0691597968139\\
    382.432807772614 232.252425130978\\
    86.9460070889919 94.604206460334\\
    113.510672872779 245.69694662296\\
    201.880330145687 72.4148501739441\\
    286.550418330002 122.399541432664\\
    155.216444378544 72.8126451750732\\
    95.6420935933005 37.1935764632019\\
    43.7635078335632 46.8277442527563\\
    169.702333681121 399.959464483236\\
    196.71505386847 298.349136576391\\
    143.289259081171 170.067880843023\\
    231.848157133439 450.223485434938\\
    372.879466266974 436.615634210823\\
    495.530129255084 418.699693052606\\
    145.995921968805 195.705688347322\\
    20.0178793505789 270.670312850296\\
    220.240786978621 412.412492161755\\
    401.449714992766 471.869000731069\\
    22.8022308669722 249.807180285836\\
    219.077842038845 364.865498827862\\
    378.451681832498 187.708839032451\\
    395.272973355523 123.553939466177\\
    159.648571904566 139.688569348141\\
    316.925728352006 201.820653767388\\
    213.171968092848 420.28751441775\\
    250.013989587867 247.742339393708\\
    463.399573670584 403.261109371915\\
    431.623770980222 25.8865640632229\\
    264.717872405636 51.4719774568903\\
    380.382441763846 74.3059179906413\\
    370.196626593404 190.641492662777\\
    486.246452255693 426.885513864484\\
    118.574626812929 251.575404496153\\
    236.905716586524 496.601907103563\\
    131.807324874403 113.294657632125\\
    361.10644214354 496.711112564119\\
    32.0645133296292 452.323699454348\\
    441.044449138916 467.989447146771\\
    222.350456070483 294.327202934972\\
    359.193542927796 38.8644324577833\\
    173.430298117729 320.842536925968\\
    167.309533199811 75.9601526436878\\
    223.073241344489 498.215035721507\\
    39.3357632218191 154.002328384122\\
    172.485602465089 453.33589883089\\
    376.81611754878 219.879349383489\\
    108.089336913481 478.147479876945\\
    339.194291603581 334.263639167072\\
    245.072818446876 312.794593719767\\
    470.117810543744 149.491376018955\\
    91.0787934701759 52.8599341869865\\
    423.481312128454 87.583385475382\\
    106.358643968349 34.4413400331133\\
    251.969705128934 348.422163465008\\
    15.5187311082854 152.47842406042\\
    337.712284894278 247.838836583482\\
    262.343762024394 130.488096560287\\
    485.071305085847 394.089191962577\\
    2.69416442663356 366.416101668102\\
    115.564957805728 202.322125887667\\
    302.093576869458 326.678439467433\\
    323.032061558841 405.763700572093\\
    435.778934363666 382.688953181166\\
    280.780438043333 223.441120467756\\
    60.9398848492882 405.806571948035\\
    382.26474981984 326.509510206771\\
    230.726368512423 14.0774349534112\\
    447.633635549916 88.525154554504\\
    258.743340299106 418.760997152078\\
    212.13028350843 149.243137901072\\
    17.1837666041285 427.363878944776\\
    317.819466348477 382.811982204452\\
    67.1675569733924 459.594098753985\\
    474.582659795531 190.590511183361\\
    157.746516656014 186.585547966193\\
    335.973398053532 449.763319528815\\
    38.0404012331303 275.077617616459\\
    279.141591126427 257.152354680414\\
    176.514490406748 61.0895270229163\\
    390.821914731648 200.494152419899\\
    237.567045880768 126.081295908062\\
    8.59695109799635 356.262570659632\\
    395.815169778709 424.92350378808\\
    345.657817814961 436.660620605542\\
    328.639360359117 246.962344191578\\
    156.851060196652 52.7138238960823\\
    161.402232200421 88.3584948009228\\
    493.036295395613 402.994868240474\\
    33.7397737256294 78.4382086514627\\
    100.896791872893 152.546394232415\\
    100.771325864056 414.412279868009\\
    487.139068939961 170.954598109543\\
    422.378969417704 79.3590417981453\\
    364.636802869243 262.585970905209\\
    429.727413833346 270.062493514744\\
    332.962463572372 400.915866183633\\
    491.654532818337 21.2912984673336\\
    135.546630563278 292.984637656965\\
    38.9793710586409 431.414947465803\\
    270.40134866551 353.94569220712\\
    440.881507383782 334.288293548379\\
    281.858974799726 367.128121432985\\
    51.3883669958144 138.846847468774\\
    227.267641151177 358.743329517165\\
    349.782934939761 276.658870479589\\
    271.052404298198 469.558084305921\\
    349.320057421009 343.742320979716\\
    357.904812203563 41.5944051099616\\
    321.788134325719 332.952139155983\\
    343.242114107458 110.262636020468\\
    293.82690683435 294.215124870464\\
    103.79652806218 357.789348326139\\
    321.851906525791 416.026049034288\\
    159.99287655797 29.6661121378661\\
    380.782723393485 102.333570126274\\
    257.71429511015 454.908753281627\\
    126.750918533512 424.335459102543\\
    417.499765335714 424.632773733531\\
    194.10333934359 65.4111941620261\\
    296.716777321877 73.8147098749009\\
    472.750020312417 459.684631159621\\
    182.990885651195 461.786320949184\\
    134.875216010446 39.5484667920163\\
    208.545803976317 302.308102711756\\
    295.608178296512 149.354915515065\\
    225.031358663261 103.085747255558\\
    96.224315439129 317.043891041147\\
    288.340919963167 247.536157916501\\
    199.498685606338 6.50672829715454\\
    110.987268280074 412.738703656492\\
    111.264505211138 276.118600539634\\
    422.220140791961 458.906713660504\\
    98.0006703533819 468.000078528716\\
    464.198510232505 88.908126016555\\
    273.96999033422 101.497515150104\\
    268.817360757099 285.262809640024\\
    373.270667077675 111.865276076999\\
    375.478396271203 173.139688844698\\
    379.149387921702 84.1525732942449\\
    389.24231478224 396.394783199368\\
    9.4294350620216 211.5253388223\\
    312.35619961191 190.343116947184\\
    };\addlegendentry{Inactive devices}
    \addplot [draw=color1, fill=color1, mark=triangle*, only marks]
    table [row sep=\\]{%
    x  y
    222.562995160743 470.403607507765\\
    126.32192881842 263.731852207188\\
    30.3951436009746 305.009568047737\\
    163.636191361421 256.903052543716\\
    464.626127776738 7.9304098948868\\
    96.5932990458944 218.31288423258\\
    408.011475554067 227.671434031174\\
    386.241426505471 39.6445216202194\\
    210.410054134695 223.903470812025\\
    181.083151386868 380.039171663711\\
    119.687390080628 82.9077386516925\\
    441.172089974567 209.633935314103\\
    84.2337944342281 387.364705052834\\
    236.866775979748 99.0643770243239\\
    };\addlegendentry{Active devices}
    
    \legend{}
    \end{axis}
    
    \end{tikzpicture}
    

%% file: img/cell-free-map.tex
\begin{tikzpicture}

            \definecolor{color0}{rgb}{0.12156862745098,0.466666666666667,0.705882352941177}
            \definecolor{color1}{rgb}{1,0.498039215686275,0.0549019607843137}
            \definecolor{color2}{rgb}{0.172549019607843,0.627450980392157,0.172549019607843}
    
    \begin{axis}[
        tick align=outside,
        tick pos=left,
        x grid style={white!69.0196078431373!black},
        xmin=-25, xmax=525,
        xtick style={color=black},
        y grid style={white!69.0196078431373!black},
        ymin=-25, ymax=525,
        ytick style={color=black},
        width=0.8\linewidth,
        height=0.8\linewidth,
        yticklabels={,,},
        scale only axis, 
    ]
    \addplot [draw=color0, fill=color0, mark=*, only marks, opacity=0.1]
    table[row sep=\\]{%
    x  y
    64.9384630931073 454.730875844874\\
    245.662760779903 120.863634457726\\
    318.044675595344 144.070822865166\\
    85.4331419728776 482.821964054575\\
    470.439207781853 448.723650725267\\
    313.213095364414 22.1517884046958\\
    153.997613739249 147.572708981291\\
    499.287759776016 91.2043498883721\\
    245.441326338527 69.4877447784283\\
    372.527792662632 109.97642157878\\
    143.906883077292 401.224674326341\\
    150.818798909608 133.085434091911\\
    329.563619512531 77.2465903032352\\
    91.2991570463059 465.51471980842\\
    2.19849795679516 469.829361844642\\
    280.233358583632 485.513425631963\\
    472.38214464312 363.360872161183\\
    378.241079231682 365.732252866519\\
    256.014182831287 36.6340285248462\\
    388.58164680901 114.133629649203\\
    160.983825626773 441.568848608022\\
    369.891661808094 250.497765738214\\
    325.886296917112 323.906194793287\\
    191.719765394961 62.3692662608432\\
    90.8081522500862 58.1295071136498\\
    360.964196157812 365.363394588264\\
    411.371190840284 59.7350754720869\\
    472.77821008412 395.56284498658\\
    114.840932742899 372.507254113088\\
    64.3904610858189 106.933548146224\\
    486.32994628173 150.095401650491\\
    409.51219512758 154.465050483524\\
    197.807947907705 73.860108602356\\
    112.299533221101 142.686118973439\\
    122.274026902067 248.224786951449\\
    136.153778158126 20.3412919061315\\
    15.6273700194509 380.638551228761\\
    491.696396293417 395.827407058873\\
    205.374450145835 277.74724786363\\
    151.297272900449 217.852613436892\\
    483.485443016232 182.009700950796\\
    461.235444275993 366.403616540931\\
    277.823369108692 326.657696317318\\
    495.934599074243 233.08243245269\\
    181.512768997491 47.0669941620931\\
    81.0622609302099 434.752631910077\\
    295.10364152989 219.670823611993\\
    163.983624489277 370.378224422928\\
    57.7419890708681 324.706127114456\\
    251.031542270877 63.8637150998588\\
    371.939058566304 270.568531653515\\
    238.479629184072 63.3281427579969\\
    395.588559233737 23.4120762432324\\
    434.21535329231 347.741857016181\\
    50.4990533321926 106.279142140497\\
    286.177990975054 9.31266935384106\\
    487.713349001768 182.788898632879\\
    402.141374808115 363.558064621095\\
    247.38707239626 53.2773852054858\\
    263.338147938154 74.3328605706963\\
    33.0654903495625 464.963903330382\\
    478.607439907609 27.2653990087082\\
    332.158857532779 137.167658905608\\
    124.02236550571 225.746479446037\\
    18.5335457439736 50.1535336122066\\
    263.774550115159 256.35972728377\\
    365.935883681716 379.978447792752\\
    303.933399936026 463.292469636297\\
    58.0531946073926 285.62104691036\\
    30.4231221696135 411.972291245246\\
    439.886321326354 219.831310031148\\
    95.2235792876249 437.445724586301\\
    123.24050292189 89.3680619376633\\
    441.754516166309 241.257048066147\\
    468.659886172886 241.353735056472\\
    417.062478869489 255.765100913287\\
    139.313000385597 473.187285916074\\
    367.791409131051 18.3175462753328\\
    497.652826374767 47.0218867806086\\
    459.029242397728 443.527550809082\\
    412.675271962832 375.021684767399\\
    463.956050243738 406.336317326663\\
    376.026747406053 97.346835616713\\
    278.828374453608 307.512280216105\\
    441.962802360933 200.257770378306\\
    422.267070401602 33.9369806925682\\
    357.866351230504 102.145941581931\\
    129.299001495282 83.5192166802122\\
    404.409672546479 228.479477095805\\
    182.368982855973 73.1612461862952\\
    381.388664408206 391.886607187981\\
    91.0948514162951 295.815822616956\\
    355.430358089875 481.738014961433\\
    274.688176190992 60.680238417199\\
    417.562789680489 256.117841636017\\
    273.25107901337 398.935191213885\\
    26.5283742638234 462.144333287043\\
    144.39249629801 55.940264772001\\
    355.701251528573 405.369792524079\\
    454.427174024094 0.272446963420658\\
    37.9140219265426 100.699650012338\\
    484.941112512727 58.6064198063263\\
    112.554173322132 368.342891019078\\
    86.1950825307584 308.235717178149\\
    75.7886616322613 466.962407372968\\
    447.208497263062 162.104593774472\\
    299.245685082418 59.5623733908272\\
    330.8898267659 391.721398108864\\
    346.977934259957 402.2736554693\\
    349.623967540541 164.708445172172\\
    195.380141259444 320.446329068191\\
    370.596837101341 480.051192156681\\
    358.947194333562 431.029516442483\\
    469.49064019035 76.1828950066023\\
    147.855708840668 154.05950687034\\
    270.536158014355 282.795000693726\\
    99.2910232227179 392.78250585021\\
    393.274482036307 72.654934848618\\
    332.094921135013 368.602853387657\\
    476.431210517195 265.337642728876\\
    392.47879082254 476.14139952637\\
    212.001260542258 463.88708944035\\
    215.985012678597 89.6443651010327\\
    370.049716318686 410.182658474704\\
    2.55328166167473 343.609748295597\\
    63.1613574242176 155.028517300474\\
    208.475845493589 177.185242420911\\
    116.276919633716 476.634584695593\\
    328.892389911033 305.803525391578\\
    28.9640357842683 0.378621008883029\\
    23.5503879710811 10.8347816839194\\
    382.914196082519 408.929933511548\\
    140.096957537079 478.822957576732\\
    350.17580317261 471.797288110117\\
    49.941499794323 191.1511830362\\
    373.407404180066 3.53967007879097\\
    421.726460245155 41.7965934093249\\
    284.399586223175 376.432347753157\\
    322.663123743897 266.920543992353\\
    72.5594450962321 426.462985611756\\
    459.017178423617 180.326925699742\\
    3.7537870983459 258.653278550675\\
    96.410081128033 455.654797829634\\
    127.85768445487 444.648673536594\\
    495.11842648113 365.008905961045\\
    261.213518851178 197.174587185071\\
    328.543598330698 265.910288347206\\
    88.747665269823 386.245590704275\\
    467.698572900258 444.971501987037\\
    170.553653075884 361.008572728094\\
    363.051519155093 488.262584005742\\
    286.08841505185 433.596243026601\\
    77.0372959600055 128.147424390062\\
    383.929105342975 385.151457102256\\
    142.869542056847 92.2346414395868\\
    378.461630939791 291.755978645085\\
    418.442893769882 11.4932270502738\\
    25.075887004655 262.442224600726\\
    271.859072221259 204.283100576423\\
    267.882461856167 242.928299595269\\
    63.4455641231591 235.968145554414\\
    35.5703638000995 319.186008594579\\
    27.216910060602 408.342189129246\\
    461.791098427187 316.02777102199\\
    74.0397894795248 131.776993115145\\
    321.087927908794 204.901254167196\\
    471.550567332733 229.247154897556\\
    75.4168137427073 348.153956712368\\
    56.5004259641323 126.090825249525\\
    168.64882175829 35.2565725791379\\
    33.0462609757969 429.339192724354\\
    26.7725636651915 450.915272577523\\
    459.233235539839 276.032849551205\\
    205.484045695385 426.253979148184\\
    204.130354879611 293.62629302835\\
    417.382546552569 210.403016285395\\
    45.0414634324631 220.329832965304\\
    497.618531428907 19.6763561439786\\
    204.050052812284 249.880294124428\\
    57.8247800520291 116.82392624238\\
    489.308590918627 495.395312937344\\
    154.433869250428 133.323969476472\\
    402.061612846143 493.845097264141\\
    248.704307643412 68.5383837847698\\
    34.9431799004293 443.938028354247\\
    330.964710066555 46.4662402602514\\
    283.110233870002 76.4173892381232\\
    2.89718339219053 216.783299517268\\
    149.664129519472 154.024349734127\\
    389.744392131155 345.158254225426\\
    156.006968864174 292.995611694811\\
    373.167288020707 276.806349437205\\
    351.934586183662 12.6502536713929\\
    491.632703495278 477.758807598436\\
    285.55376175013 45.2736054780359\\
    293.52444517893 280.448513459086\\
    141.480658284369 146.407313238409\\
    159.182051578613 393.004352211542\\
    0.812956167566325 104.135302692941\\
    460.774815928997 345.725295812487\\
    347.035242899397 156.903612328356\\
    418.176941593216 354.603489720881\\
    39.7527327405764 203.416755158029\\
    100.159716276543 181.77439806228\\
    221.271115292828 101.694317696418\\
    112.765250986341 228.058794309985\\
    90.9484701884102 173.069357795223\\
    429.959188158895 426.730715633196\\
    277.290614236043 487.398884394392\\
    213.217205458547 438.309439681715\\
    429.596819707505 497.098603398645\\
    58.0442967137685 151.415952680487\\
    382.866238620498 86.6882826902952\\
    205.959691244491 422.291731763618\\
    220.445522123469 8.04807968375781\\
    291.572805631621 268.991907088113\\
    76.190463998728 273.852057359903\\
    10.6583589030569 418.928602980972\\
    21.7477977205611 480.779332713886\\
    292.567929501634 240.448852709863\\
    282.355400764788 44.7932804831399\\
    222.721224141543 254.174339442859\\
    435.8292397566 386.999745703782\\
    397.302593494521 470.42469200646\\
    238.277528434523 279.544098467624\\
    300.298044834172 485.532289342294\\
    322.099003206017 468.403727595025\\
    123.766555684276 289.09857445754\\
    345.187613503981 62.6242168369271\\
    359.091150370236 426.431262038621\\
    431.611633312447 217.494283874309\\
    238.634323757128 104.007245839379\\
    191.797034969695 203.888867880197\\
    206.491785737417 220.391942302841\\
    162.614988063084 324.588373820955\\
    499.154822018112 445.185319642341\\
    150.010794259308 189.15909839878\\
    76.2823523178276 176.002085766477\\
    102.602758068251 127.008387164458\\
    396.70676626766 458.483147522455\\
    158.983750619556 99.3027468911721\\
    447.547655492994 335.915485674062\\
    151.806039668577 148.376396557313\\
    351.469146313992 193.727120299708\\
    203.246868806576 1.90827414429962\\
    244.196066647327 271.471127976884\\
    136.907857268178 14.4488186167075\\
    422.305681945899 45.3586781575613\\
    235.907015174207 90.0913195936377\\
    39.8732043751722 163.533549005172\\
    15.865220102561 164.146402777546\\
    250.747501102713 181.603699484528\\
    51.473368096855 284.856923367219\\
    168.659661635049 362.660873786492\\
    272.905769037288 411.499452765468\\
    463.47519111877 222.562995160743\\
    470.403607507765 126.32192881842\\
    263.731852207188 201.506378558326\\
    471.372359742329 298.519415444749\\
    333.11222360294 455.55389114742\\
    211.456260178687 401.116196434727\\
    169.322775647102 437.362990909153\\
    466.432100879868 40.9722587348114\\
    485.694390324735 123.370094007119\\
    117.26349816936 87.8256403634233\\
    460.081860845924 325.623860065416\\
    335.81776726045 318.175304296699\\
    397.028402903851 123.803565875205\\
    187.099151783001 254.763792270727\\
    298.485788995258 363.097336680844\\
    141.039790668592 392.558650168972\\
    270.436530670064 249.716116439744\\
    448.165812411041 459.632173336458\\
    307.65532419955 126.63457382109\\
    247.114916604083 1.13663122321006\\
    334.229074361364 192.574782271769\\
    245.9699999071 498.309149873147\\
    245.738233521224 273.415485901719\\
    499.540727213266 144.757874277679\\
    226.862343874721 10.4228711799573\\
    309.442128894789 54.662197348671\\
    427.042267044867 415.220141440814\\
    297.292594239848 100.961291272675\\
    152.064464513022 336.378405014999\\
    78.942414228637 336.962201470663\\
    417.493472007369 27.86825992967\\
    179.537143408907 22.7969237668469\\
    24.9681362673652 58.7883264922893\\
    19.2823542765266 318.718264972533\\
    133.321734577247 346.618129949346\\
    390.0188783495 86.3426753014914\\
    222.735917572245 352.958150448585\\
    75.9179485152801 152.320063032784\\
    3.09894020821583 71.4122623182641\\
    335.896920526867 358.930023927155\\
    410.158906009803 323.507300749364\\
    435.154118579919 428.15030768338\\
    19.158689255965 45.1738608665622\\
    90.3927346011665 464.268147156035\\
    185.788589702156 256.620728698177\\
    158.84770230259 416.335997227721\\
    136.062879907615 240.941144661804\\
    263.280149119155 384.47024000595\\
    60.1170921631858 120.904209863949\\
    250.652281121505 86.559063606799\\
    460.142119926363 423.311711741138\\
    86.9238576254883 2.60376640679311\\
    153.832969555133 136.993236188047\\
    309.386051500792 403.387082698616\\
    42.77135339968 349.922741102947\\
    90.4230996917377 121.191071114406\\
    485.152279650117 60.5094913857941\\
    54.2446543759556 481.256848335314\\
    138.538286317653 328.690095057579\\
    216.940602924311 472.505453126243\\
    146.758760405178 95.8513962772983\\
    466.605190297031 292.943106748094\\
    387.661672254777 293.32894460275\\
    454.387961016807 36.1899134978801\\
    223.845541479839 430.449526163641\\
    270.746043353223 336.932910342125\\
    447.845791526155 380.069132074195\\
    226.610911046093 466.99726915268\\
    208.192757080887 111.056529052537\\
    115.251176828181 62.9973232463695\\
    11.555875037406 27.2891774977888\\
    183.22723954364 401.070075362191\\
    286.898866195256 359.080881225841\\
    118.286188068486 99.0796537727492\\
    484.685247334006 417.393900099844\\
    411.454747298179 204.486020026141\\
    227.939073958929 304.241169990389\\
    184.73209129827 60.96203838895\\
    249.703642206093 331.468175571561\\
    28.2581872425065 499.9456775868\\
    134.900251112982 264.229310081275\\
    377.047807834306 113.552549434178\\
    314.149892122409 421.373709315519\\
    255.518697708452 378.728785535828\\
    365.353973877083 351.306768284033\\
    304.830141321346 277.196813900792\\
    253.865620305548 465.470691915917\\
    322.037173491556 95.5203921720639\\
    403.524024709267 77.4458051799253\\
    288.981433966368 464.102793312517\\
    316.638134430004 74.193854807392\\
    241.207465173113 418.992840220393\\
    193.357708326874 184.019689556145\\
    302.315071927867 228.19605195384\\
    9.57071572396462 472.396844435735\\
    227.044949068707 328.758488197195\\
    492.079058681402 345.883692438664\\
    257.768500238076 402.747572545138\\
    95.7926955244308 495.782286718262\\
    480.837931848851 148.542500031413\\
    301.437973401857 258.209290784321\\
    100.268576024666 313.895723848676\\
    387.135359016927 63.0631922365245\\
    42.8387321612512 397.367533301081\\
    283.912005558171 287.74196870496\\
    71.5740298890724 200.049129025739\\
    326.670855540801 112.518432171703\\
    429.731780729899 107.520292754536\\
    410.496163746502 57.2443190065537\\
    42.4496926601553 350.274663301403\\
    85.7795600163211 200.360767367044\\
    231.453075794505 447.084709961588\\
    74.7306247916925 191.999001395543\\
    458.11093183166 248.359575447356\\
    163.230970289465 20.5042684212365\\
    363.246818664278 248.718745732034\\
    136.565195667762 313.729514257329\\
    40.0879158707818 97.4454799227119\\
    317.08966541498 83.9966193726703\\
    499.544916870037 98.4275271947832\\
    432.565300025795 428.726886343129\\
    389.937160884782 431.356370675505\\
    349.831657198785 151.700857921428\\
    432.894259591963 254.840537180012\\
    399.950982579525 470.534250096678\\
    101.036847163621 427.412745623024\\
    175.280666340019 127.179683588962\\
    468.782315103552 214.915150283908\\
    328.924202645531 356.735143677256\\
    166.525896790244 467.076401120849\\
    271.729234738076 151.537127677041\\
    206.627642787837 112.789686108012\\
    399.917612079894 410.788797401929\\
    277.289473140683 197.495808370415\\
    196.714564707061 279.28438257969\\
    417.203146188078 343.681070774048\\
    189.208829442948 97.0940878577286\\
    230.096117433547 335.211698176753\\
    223.002451931118 410.336889433903\\
    141.598713285334 476.10128903218\\
    290.772765779076 84.6255198746449\\
    398.156011592846 374.17781473239\\
    332.067713197384 199.455207017176\\
    186.458114601936 63.5797941745555\\
    304.224676296139 469.816350827095\\
    447.394964520419 368.061612103459\\
    239.584082932599 433.280486053445\\
    472.777180151736 207.016568895874\\
    440.369208010751 293.666682651537\\
    365.218344747278 364.160833590462\\
    375.325833648317 262.192835754898\\
    163.447987871005 380.975980096996\\
    448.442391813999 391.632339513928\\
    176.952155703322 13.2327337233236\\
    4.08086901184429 283.297038639038\\
    123.170897944401 205.065319025253\\
    133.945984926157 91.4843192779597\\
    0.821207278592695 225.481399429388\\
    99.7380742423415 255.66217653068\\
    479.024481053067 325.946839568393\\
    237.985879261108 72.4077124643764\\
    484.462111078597 384.535039812078\\
    164.43668664712 412.373244266754\\
    393.231486607253 272.0057175752\\
    258.678216759218 166.137084777068\\
    176.729281607413 306.089276904843\\
    295.276392422497 169.063332673594\\
    112.706863591958 336.606946143349\\
    125.076866646639 477.037354750096\\
    407.434644805357 313.281540222178\\
    362.463738976112 436.279313342579\\
    298.136202524883 321.198273597564\\
    498.961508384998 499.600138575464\\
    104.006002570057 240.561038691837\\
    295.16742639337 163.636191361421\\
    256.903052543716 76.494837853682\\
    139.873368862426 494.131536333844\\
    89.9924487856304 464.626127776738\\
    7.9304098948868 86.0090606833642\\
    1.86293588441855 357.651546439624\\
    279.837829092479 96.5932990458944\\
    218.31288423258 420.407602525836\\
    363.918809844994 484.934583940738\\
    358.01105077694 249.239494478512\\
    62.6126423691801 199.430197851778\\
    373.216351306942 104.93177026085\\
    490.865133056387 20.7746612378166\\
    5.43089172694622 338.012366398729\\
    395.761421961836 97.506051142187\\
    190.795620108131 396.497989714963\\
    314.072702730295 71.086349239718\\
    14.6607330585214 431.853977555933\\
    364.333662228565 460.677187090644\\
    483.743575432131 34.6172395449261\\
    358.804847663158 423.759025468613\\
    126.801127279156 247.685911494976\\
    22.6773071707206 372.782019247487\\
    241.638631797094 197.247860123563\\
    329.552474855122 428.096141734913\\
    289.011954625744 18.4236852725387\\
    311.315457431876 49.3162649001074\\
    43.0320490867326 137.306459177901\\
    252.015776673926 33.522042540286\\
    25.7815590043991 382.542012115407\\
    68.350900716766 213.371003615289\\
    20.670627398315 182.245919698751\\
    29.2769892741873 482.9407331817\\
    227.722370699333 167.39030326544\\
    292.096066702021 284.271335754327\\
    215.682354555882 450.857976601719\\
    445.813388247692 152.361252684243\\
    364.448157413336 394.312188006776\\
    4.96533561165491 272.310585848313\\
    418.780994371999 254.146522900541\\
    435.122150505596 92.6567132943707\\
    177.281255842727 91.0794262769919\\
    178.733046859993 257.453112856832\\
    33.1038238956787 62.7548056569631\\
    251.39122824241 91.5039219399045\\
    169.635961682734 234.460527695316\\
    186.378533990824 260.888795844826\\
    352.751142808158 206.716728598542\\
    49.8216844323334 9.42963636693922\\
    218.676843062069 168.479690546406\\
    18.3934415533596 99.4749825863598\\
    24.4028881411173 469.03722362976\\
    269.114343342971 313.526761855946\\
    384.672080582989 345.628546122784\\
    93.149816829786 152.520236501768\\
    325.790056069292 132.529985675338\\
    91.8294834216131 410.873815587418\\
    320.631115717882 168.039906408767\\
    458.448255578457 302.993744278228\\
    310.461150035517 170.775686693078\\
    23.8748050676239 304.736415167667\\
    297.131738811075 122.55563776323\\
    440.516180669231 129.883588969131\\
    290.321786358056 385.510022925188\\
    351.124892603358 129.288109876252\\
    123.22075059756 70.6220907156006\\
    459.684421544316 223.616819717394\\
    20.8867397219618 290.358160425711\\
    466.254099100954 339.290132866772\\
    440.419294316937 454.457432359202\\
    446.294644664717 278.901356981207\\
    357.990900375717 460.805971364807\\
    439.476949622451 86.9406888683164\\
    438.889341131468 322.321274712481\\
    16.9113674187008 154.293453360174\\
    420.93622736633 441.581482265659\\
    38.3062189051349 443.579132600868\\
    129.230121056561 12.5189668207118\\
    221.358957870478 415.274552842847\\
    248.670525176238 475.338234826965\\
    300.202117176352 387.077335016815\\
    382.96001682567 195.580382229661\\
    173.837057336554 282.758893030291\\
    167.315845396248 115.910210441828\\
    32.0672435416255 265.322178731347\\
    218.635858883949 376.627439228509\\
    41.6385403197799 209.666122229001\\
    270.886075634049 394.270965469065\\
    155.941999099846 158.671294199798\\
    9.31110345686331 124.87053511956\\
    44.7396853714928 311.216468288144\\
    73.4924412301661 249.417128990457\\
    417.018827779818 483.657143572591\\
    373.198798562666 457.901084868955\\
    397.245953312734 428.715258940648\\
    257.253779782766 207.489641466073\\
    248.151190015492 156.327424451711\\
    85.8143474705125 77.9129477618715\\
    396.68114372921 45.4070282614247\\
    56.3319656536861 89.081422725878\\
    413.394731573902 302.883576869806\\
    425.965080688404 30.9020137899043\\
    440.338254794455 255.826300443825\\
    33.3790113153648 74.1423901547801\\
    410.244258685477 302.834949442203\\
    438.052353918077 476.42870097441\\
    430.074418867574 416.567571235842\\
    264.984662333554 366.100659540948\\
    353.569111202935 261.655255678217\\
    327.261330579807 394.85651334715\\
    441.453527135054 308.790042942292\\
    277.164373263785 323.166436766421\\
    444.753614818009 84.7427750738616\\
    378.837991327281 403.144950206051\\
    136.638040161647 111.428974918088\\
    245.654231837422 55.2954646360655\\
    472.498759728045 112.596956400873\\
    277.99181625717 253.936680269003\\
    364.429439633529 408.011475554067\\
    227.671434031174 116.977387414029\\
    254.115454471435 116.702991998787\\
    461.218224266087 364.557890691295\\
    38.8196676610633 146.613442478245\\
    99.7390496883767 114.155283226648\\
    369.953313758736 118.976885794845\\
    273.593582104944 170.273421907271\\
    449.364031972215 281.020420335399\\
    104.779529912708 354.51567268505\\
    225.753149313624 206.272895814716\\
    88.4572134639826 429.147294242418\\
    276.122117633375 301.237318005074\\
    22.4951383749689 442.516990042866\\
    236.090317726719 160.549247704769\\
    355.896376121005 10.6618167733585\\
    257.436733434655 388.144299331249\\
    297.947308569313 371.622431046919\\
    150.172523661899 465.319231932176\\
    190.065918386169 164.878623373057\\
    398.040156295367 460.376780965122\\
    247.111793937615 50.3308590054002\\
    38.2366309994845 94.9186392119644\\
    159.878133910178 66.3398005831694\\
    490.058575739707 49.1576749559245\\
    357.290285160844 17.9240894557608\\
    37.4540530988633 219.327092214832\\
    163.789477291217 377.185085110356\\
    329.418873145238 494.528033504284\\
    198.473720565433 115.086971707281\\
    359.926507146286 409.462545815263\\
    14.5222145761144 6.79376421320793\\
    329.001655369942 88.8513885334264\\
    403.574198413886 234.958149003508\\
    417.912005803844 51.5690942511067\\
    286.668178617932 407.748890553167\\
    362.692035241822 239.72004990793\\
    116.194113296253 204.011161378654\\
    247.233687577587 135.258484529991\\
    69.945640210788 410.029644948416\\
    0.722548523617139 250.165335464365\\
    231.813156064149 141.894343848419\\
    282.732720494229 411.580137877465\\
    135.518261514067 259.037135761369\\
    372.83232710426 418.236819062714\\
    454.889976171566 53.1187401482863\\
    28.2170231167049 381.065563241943\\
    363.215396288029 449.87158772878\\
    41.6443640288399 223.636658211238\\
    215.82772053373 102.637678922373\\
    103.050673330147 5.75036192054829\\
    491.113776131868 238.640860493246\\
    149.975983435738 204.639160567514\\
    140.130481202486 7.17334217723853\\
    196.681682762283 70.9704084310049\\
    28.4517881989729 219.400607515389\\
    260.727737973644 142.950263551393\\
    291.820912719496 54.8127300989341\\
    224.658358160791 146.974202748304\\
    338.068457233154 466.95766445648\\
    367.617143310442 344.568928063753\\
    1.13959710041006 274.601065983703\\
    294.848932905896 243.530772726906\\
    366.380406072125 448.47582071203\\
    280.217582587976 424.695639985854\\
    52.4450390468817 357.62729734412\\
    464.937963378679 298.675673036866\\
    179.27894844702 270.822874052071\\
    109.566449858281 366.497539444872\\
    3.04016464905338 386.241426505471\\
    39.6445216202194 10.8316247381112\\
    253.97835304387 271.74856006147\\
    247.998996158568 125.221779933981\\
    192.893835415468 355.991829683548\\
    435.435133817447 425.883300936487\\
    201.392485235782 360.606422422793\\
    176.825139086723 88.36790652865\\
    413.36606144983 81.7493752883062\\
    360.761550845745 223.359417241621\\
    130.745969528354 203.076451737736\\
    246.165229649005 311.062022889012\\
    197.607579693486 136.943185393171\\
    354.604541524963 132.07350843189\\
    380.656998742825 455.331503103858\\
    407.446339969648 29.7696899143189\\
    91.2797604882228 361.952103096624\\
    171.25171242567 208.961899723023\\
    157.21871719068 199.075363756919\\
    374.882813576191 150.375862584026\\
    67.8917430458952 268.421336888703\\
    493.212035964632 397.651223360278\\
    294.459163818319 402.042762344609\\
    7.05752590612702 387.412249320192\\
    460.471590685213 484.823469045488\\
    256.265547613208 113.184354847619\\
    385.998365716677 447.020054870679\\
    384.803781029374 287.666055048804\\
    156.734067027295 141.534129867629\\
    275.652110176992 244.435697048488\\
    200.565860891158 424.955960774324\\
    212.400571345514 196.547083817694\\
    498.616107984368 228.647032369692\\
    496.032074879246 252.539939023146\\
    315.956403646606 160.671592218136\\
    29.9067900403389 140.013625209693\\
    252.257350368665 188.817383505563\\
    61.2977229589518 258.613543903207\\
    26.89213879668 499.552446760012\\
    182.254578329777 80.4342838532666\\
    81.4458469102726 9.66437985150886\\
    441.466569800765 325.413685485616\\
    470.699343837153 278.552066257535\\
    411.659297569882 452.971419661761\\
    315.185565458496 244.023099313913\\
    191.864303637413 459.476961960265\\
    45.0888289987454 16.626029315719\\
    224.977692295722 387.631545205931\\
    419.912450134257 403.922580964848\\
    113.587405488484 463.795229272965\\
    311.498857514017 148.841732778572\\
    426.680324830588 388.859300878882\\
    212.503290207789 380.319002603798\\
    149.4403668774 344.195202750616\\
    278.70021363808 353.024184123784\\
    128.315505327487 336.24433525023\\
    364.120181510889 313.96757656994\\
    130.426217773423 262.248842793892\\
    396.204693966226 59.5177451697938\\
    456.282692906734 391.995990102316\\
    50.0591763845718 378.773377728712\\
    268.065000997547 234.875136329515\\
    91.3286283654803 458.822215613317\\
    305.761533023149 214.568526217936\\
    226.760882162949 76.4938711175617\\
    413.919917740723 268.643910545002\\
    470.108802652589 253.401658492631\\
    411.747732095327 395.209757618334\\
    438.146119444567 351.088211054831\\
    62.9465896406791 223.436042898553\\
    83.3349090303822 11.5355892478439\\
    44.4029725184532 319.040477738471\\
    375.558580649971 361.035753051443\\
    248.75047315269 389.928524822231\\
    85.2780016174208 451.980221166012\\
    378.68250794405 257.609425871833\\
    243.859890886924 324.556846899068\\
    43.7624157195775 269.84661459407\\
    215.204146383572 182.838322153892\\
    361.364062005838 132.555701854065\\
    349.935401936275 146.96681935343\\
    383.481791814396 10.2183039998891\\
    213.444030080857 364.599257878207\\
    266.606204232984 255.434236015291\\
    133.735187574397 260.245564694593\\
    207.981610487125 148.419836333394\\
    215.019655196399 143.706999456785\\
    239.600658412893 282.76685036663\\
    378.067934212063 231.38614850165\\
    279.144209096244 11.8094849843322\\
    25.5956484806581 378.604814044552\\
    68.1594879629358 302.6993185789\\
    358.453802483438 18.7077311083565\\
    11.6346135640826 332.391594263717\\
    117.957483071052 328.725205239196\\
    278.78780284751 75.4056183099671\\
    21.5735046815225 146.550544917977\\
    336.79590009174 360.233741189262\\
    471.901386882779 356.697544407565\\
    422.2386606351 156.649356644395\\
    110.529046551063 118.928587430975\\
    484.377647862901 64.9549020101737\\
    182.055527131071 258.588505039456\\
    165.450636670678 89.8864481781199\\
    405.88972359045 30.9825868812748\\
    377.832284635691 262.52726866832\\
    156.655583951783 348.51998857475\\
    436.095316271804 210.410054134695\\
    223.903470812025 199.267662514411\\
    248.642435025004 308.3842769033\\
    408.164586934088 201.823513011845\\
    182.451610366237 4.20580537488707\\
    324.90005361136 365.071664733413\\
    107.646022789712 355.311478458005\\
    488.823172261508 248.501753964084\\
    209.91559530341 189.396349008439\\
    34.3705919422284 66.4187753914961\\
    422.203177411943 431.229706589471\\
    228.534265498074 481.383362857083\\
    67.1394076841864 1.74111499231372\\
    334.067988173499 296.439806895895\\
    361.52008324006 146.971839920332\\
    410.925221555424 179.278200539604\\
    490.379372156643 431.234257761026\\
    99.1739720863137 484.234023515501\\
    366.124500596844 25.7030584422015\\
    112.129242048795 89.6426361606763\\
    331.973581311632 433.552084105249\\
    11.8897544999703 395.509345611328\\
    59.3531830411618 422.278489755848\\
    59.8268870894273 420.174237409198\\
    431.176754099119 382.935230855501\\
    302.913678088358 133.259937654625\\
    390.398426948387 130.539144885867\\
    328.242269703339 65.6868473097534\\
    297.66314963778 406.942863516031\\
    76.0357398098415 121.821506875278\\
    66.0178228994483 137.789416425434\\
    142.105605377501 221.056058895422\\
    77.3393152389458 293.435110545292\\
    178.309771594326 231.032731870165\\
    69.3924382937395 361.765742503936\\
    458.993586633006 357.158561087919\\
    230.430899041971 341.977763892559\\
    128.870777437342 40.0425396468054\\
    280.450115962617 428.271402469684\\
    162.47393548204 167.327857325078\\
    87.3412735647954 111.347951824388\\
    487.438226933504 497.052942091855\\
    136.529029332451 61.5651378755996\\
    315.436341908862 445.69953244009\\
    31.8890126016262 289.792272739318\\
    132.332570150122 490.214653164036\\
    216.895359875399 40.5315225942229\\
    87.5655908198234 319.858872744572\\
    272.615303962327 377.050354402829\\
    368.811077276193 212.168566888836\\
    422.691536334577 431.62772178195\\
    164.262071273693 371.791056656899\\
    43.8951428803275 272.999439094579\\
    417.629032528677 399.44276198582\\
    320.743494182871 253.359519133383\\
    342.151424769678 103.571598841205\\
    10.9349048349317 496.056582564567\\
    105.8467288613 55.7409728386087\\
    481.446377724931 368.238585947091\\
    79.3159831938782 107.199400420626\\
    150.291123753881 304.234559343172\\
    376.66247888607 136.252981957482\\
    95.8657880809406 283.977550818831\\
    4.74501586918113 325.077620013174\\
    409.74810121406 181.083151386868\\
    380.039171663711 372.903947365976\\
    218.56725783618 498.276072341265\\
    101.61901100486 396.083237451207\\
    122.511711178612 87.773558397534\\
    327.721524814276 86.8807673501469\\
    278.544973344374 63.4458653788014\\
    425.54470296794 40.8465914868966\\
    222.051021743359 278.234234903249\\
    325.77367069565 420.35345751118\\
    223.230458122302 79.1443774099709\\
    450.181471072113 123.469777799902\\
    291.451347634124 398.648410248982\\
    389.240949554592 163.284545607035\\
    205.957507204285 95.801810903006\\
    291.462181614896 343.372514955166\\
    180.230707985475 199.221043251758\\
    13.6727315816791 75.6275919703936\\
    496.810978277317 246.004223992852\\
    398.522641150814 51.9519190265165\\
    268.66066789364 394.640541653542\\
    182.176788688616 105.522889830389\\
    315.465428809941 282.552050016749\\
    394.251379786259 383.271278143609\\
    453.58622198514 419.921486112785\\
    241.628575589159 407.959239451748\\
    32.4060845833857 4.70580820352806\\
    364.703410509919 495.501978885698\\
    353.545629768281 318.696069213317\\
    381.45571678444 326.553116305388\\
    404.539929948942 100.689503468991\\
    439.014090594459 463.103613637543\\
    162.473153663017 442.164015595153\\
    386.143552596075 276.66490072406\\
    381.57414305058 491.411611608058\\
    132.940375498561 422.502304844671\\
    425.492533525794 487.573208273725\\
    274.980509871934 451.417207770078\\
    51.6373036683337 395.080986881901\\
    175.838284940426 269.588193222371\\
    295.798193066261 459.030451007108\\
    317.6860064187 154.874425823668\\
    285.864407587977 9.70744857533973\\
    396.789458470347 357.304879012271\\
    123.642993690618 72.2798513267278\\
    344.908286399603 453.335335458946\\
    94.0691597968139 382.432807772614\\
    232.252425130978 86.9460070889919\\
    94.604206460334 113.510672872779\\
    245.69694662296 201.880330145687\\
    72.4148501739441 286.550418330002\\
    122.399541432664 155.216444378544\\
    72.8126451750732 119.687390080628\\
    82.9077386516925 95.6420935933005\\
    37.1935764632019 43.7635078335632\\
    46.8277442527563 169.702333681121\\
    399.959464483236 196.71505386847\\
    298.349136576391 143.289259081171\\
    170.067880843023 231.848157133439\\
    450.223485434938 372.879466266974\\
    436.615634210823 495.530129255084\\
    418.699693052606 145.995921968805\\
    195.705688347322 20.0178793505789\\
    412.412492161755 401.449714992766\\
    471.869000731069 22.8022308669722\\
    249.807180285836 219.077842038845\\
    364.865498827862 378.451681832498\\
    187.708839032451 395.272973355523\\
    123.553939466177 159.648571904566\\
    139.688569348141 316.925728352006\\
    201.820653767388 213.171968092848\\
    420.28751441775 250.013989587867\\
    247.742339393708 463.399573670584\\
    403.261109371915 431.623770980222\\
    25.8865640632229 264.717872405636\\
    51.4719774568903 380.382441763846\\
    74.3059179906413 370.196626593404\\
    190.641492662777 486.246452255693\\
    426.885513864484 118.574626812929\\
    251.575404496153 236.905716586524\\
    496.601907103563 131.807324874403\\
    113.294657632125 361.10644214354\\
    496.711112564119 32.0645133296292\\
    452.323699454348 441.044449138916\\
    467.989447146771 222.350456070483\\
    294.327202934972 359.193542927796\\
    38.8644324577833 173.430298117729\\
    320.842536925968 167.309533199811\\
    75.9601526436878 223.073241344489\\
    498.215035721507 39.3357632218191\\
    154.002328384122 172.485602465089\\
    453.33589883089 376.81611754878\\
    219.879349383489 108.089336913481\\
    478.147479876945 339.194291603581\\
    334.263639167072 245.072818446876\\
    312.794593719767 470.117810543744\\
    149.491376018955 91.0787934701759\\
    52.8599341869865 423.481312128454\\
    87.583385475382 106.358643968349\\
    34.4413400331133 251.969705128934\\
    348.422163465008 15.5187311082854\\
    152.47842406042 337.712284894278\\
    247.838836583482 262.343762024394\\
    130.488096560287 485.071305085847\\
    394.089191962577 2.69416442663356\\
    366.416101668102 115.564957805728\\
    202.322125887667 302.093576869458\\
    326.678439467433 323.032061558841\\
    405.763700572093 435.778934363666\\
    382.688953181166 280.780438043333\\
    223.441120467756 60.9398848492882\\
    405.806571948035 382.26474981984\\
    326.509510206771 230.726368512423\\
    14.0774349534112 447.633635549916\\
    88.525154554504 258.743340299106\\
    418.760997152078 212.13028350843\\
    149.243137901072 17.1837666041285\\
    427.363878944776 317.819466348477\\
    382.811982204452 67.1675569733924\\
    459.594098753985 474.582659795531\\
    190.590511183361 157.746516656014\\
    186.585547966193 335.973398053532\\
    449.763319528815 441.172089974567\\
    209.633935314103 38.0404012331303\\
    275.077617616459 279.141591126427\\
    257.152354680414 84.2337944342281\\
    387.364705052834 176.514490406748\\
    61.0895270229163 390.821914731648\\
    200.494152419899 237.567045880768\\
    126.081295908062 8.59695109799635\\
    356.262570659632 395.815169778709\\
    424.92350378808 345.657817814961\\
    436.660620605542 328.639360359117\\
    52.7138238960823 161.402232200421\\
    88.3584948009228 493.036295395613\\
    78.4382086514627 100.896791872893\\
    152.546394232415 100.771325864056\\
    414.412279868009 487.139068939961\\
    170.954598109543 422.378969417704\\
    79.3590417981453 364.636802869243\\
    262.585970905209 429.727413833346\\
    270.062493514744 332.962463572372\\
    400.915866183633 491.654532818337\\
    21.2912984673336 135.546630563278\\
    292.984637656965 38.9793710586409\\
    431.414947465803 270.40134866551\\
    353.94569220712 440.881507383782\\
    334.288293548379 281.858974799726\\
    367.128121432985 51.3883669958144\\
    138.846847468774 227.267641151177\\
    358.743329517165 349.782934939761\\
    276.658870479589 271.052404298198\\
    469.558084305921 349.320057421009\\
    343.742320979716 357.904812203563\\
    41.5944051099616 321.788134325719\\
    332.952139155983 343.242114107458\\
    110.262636020468 293.82690683435\\
    294.215124870464 103.79652806218\\
    357.789348326139 321.851906525791\\
    416.026049034288 159.99287655797\\
    29.6661121378661 380.782723393485\\
    102.333570126274 257.71429511015\\
    454.908753281627 126.750918533512\\
    424.335459102543 417.499765335714\\
    424.632773733531 194.10333934359\\
    65.4111941620261 296.716777321877\\
    73.8147098749009 472.750020312417\\
    459.684631159621 182.990885651195\\
    461.786320949184 134.875216010446\\
    39.5484667920163 208.545803976317\\
    302.308102711756 295.608178296512\\
    149.354915515065 236.866775979748\\
    99.0643770243239 225.031358663261\\
    103.085747255558 96.224315439129\\
    317.043891041147 288.340919963167\\
    247.536157916501 199.498685606338\\
    6.50672829715454 110.987268280074\\
    412.738703656492 111.264505211138\\
    276.118600539634 422.220140791961\\
    458.906713660504 98.0006703533819\\
    468.000078528716 464.198510232505\\
    88.908126016555 273.96999033422\\
    285.262809640024 373.270667077675\\
    111.865276076999 375.478396271203\\
    173.139688844698 379.149387921702\\
    84.1525732942449 389.24231478224\\
    396.394783199368 9.4294350620216\\
    211.5253388223 312.35619961191\\
    190.343116947184 393.505564864758\\
    371.204495609494 19.9234458030636\\
    184.052585581036 369.794842080365\\
    351.57827606155 320.275778801292\\
    310.582257563313 153.794232682644\\
    309.268274094646 276.816187361497\\
    487.151534387754 267.996781037794\\
    31.4147865698033 23.2929598527231\\
    357.615443339044 329.197562329389\\
    84.3612247864686 3.22832687376445\\
    498.127036314728 248.586523428175\\
    };
    \addplot [draw=color1, fill=color1, mark=triangle*, only marks]
    table[row sep=\\]{%
    x  y
    240.871731314639 30.3951436009746\\
    305.009568047737 14.5446308899832\\
    35.7359479046061 407.163350924516\\
    402.529289141575 189.275503876651\\
    461.618587889119 355.447080861458\\
    166.622980758624 191.604431670783\\
    260.480721224011 299.065578719556\\
    227.192695853279 333.019348317558\\
    14.2665438610435 109.151349356028\\
    417.50595072921 403.919781933584\\
    270.670312850296 220.240786978621\\
    246.962344191578 156.851060196652\\
    402.994868240474 33.7397737256294\\
    101.497515150104 268.817360757099\\
    };
    \addplot [draw=color2, fill=color2, mark=square*, only marks, mark size=1.5pt]
    table[row sep=\\]{%
    x  y
    428.571428571429 142.857142857143\\
    428.571428571429 285.714285714286\\
    428.571428571429 71.4285714285714\\
    357.142857142857 428.571428571429\\
    214.285714285714 428.571428571429\\
    285.714285714286 71.4285714285714\\
    428.571428571429 428.571428571429\\
    285.714285714286 142.857142857143\\
    428.571428571429 357.142857142857\\
    214.285714285714 71.4285714285714\\
    357.142857142857 357.142857142857\\
    357.142857142857 142.857142857143\\
    357.142857142857 71.4285714285714\\
    71.4285714285714 71.4285714285714\\
    285.714285714286 428.571428571429\\
    285.714285714286 357.142857142857\\
    71.4285714285714 357.142857142857\\
    142.857142857143 285.714285714286\\
    428.571428571429 214.285714285714\\
    357.142857142857 214.285714285714\\
    142.857142857143 428.571428571429\\
    214.285714285714 142.857142857143\\
    142.857142857143 357.142857142857\\
    214.285714285714 357.142857142857\\
    214.285714285714 285.714285714286\\
    285.714285714286 214.285714285714\\
    142.857142857143 142.857142857143\\
    357.142857142857 285.714285714286\\
    71.4285714285714 428.571428571429\\
    142.857142857143 214.285714285714\\
    71.4285714285714 285.714285714286\\
    214.285714285714 214.285714285714\\
    };
    \end{axis}
    
    \end{tikzpicture}
    

%% file: conference_041818.bbl
\begin{thebibliography}{25}
\providecommand{\natexlab}[1]{#1}
\providecommand{\url}[1]{#1}
\csname url@samestyle\endcsname
\providecommand{\newblock}{\relax}
\providecommand{\bibinfo}[2]{#2}
\providecommand{\BIBentrySTDinterwordspacing}{\spaceskip=0pt\relax}
\providecommand{\BIBentryALTinterwordstretchfactor}{4}
\providecommand{\BIBentryALTinterwordspacing}{\spaceskip=\fontdimen2\font plus
\BIBentryALTinterwordstretchfactor\fontdimen3\font minus
  \fontdimen4\font\relax}
\providecommand{\BIBforeignlanguage}[2]{{%
\expandafter\ifx\csname l@#1\endcsname\relax
\typeout{** WARNING: IEEEtranN.bst: No hyphenation pattern has been}%
\typeout{** loaded for the language `#1'. Using the pattern for}%
\typeout{** the default language instead.}%
\else
\language=\csname l@#1\endcsname
\fi
#2}}
\providecommand{\BIBdecl}{\relax}
\BIBdecl

\bibitem[Tuset-Peiro et~al.(2015)Tuset-Peiro, Vazquez-Gallego, Alonso-Zarate,
  Alonso, and Vilajosana]{TUSETPEIRO201584}
\BIBentryALTinterwordspacing
P.~Tuset-Peiro, F.~Vazquez-Gallego, J.~Alonso-Zarate, L.~Alonso, and
  X.~Vilajosana, ``{LPDQ: A self-scheduled TDMA MAC protocol for one-hop
  dynamic low-power wireless networks},'' \emph{Pervasive and Mobile
  Computing}, vol.~20, pp. 84--99, 2015. [Online]. Available:
  \url{https://www.sciencedirect.com/science/article/pii/S1574119214001576}
\BIBentrySTDinterwordspacing

\bibitem[Wang et~al.(2015)Wang, Dai, Yuan, and Wang]{7390876}
B.~Wang, L.~Dai, Y.~Yuan, and Z.~Wang, ``{Compressive Sensing Based Multi-User
  Detection for Uplink Grant-Free Non-Orthogonal Multiple Access},'' in
  \emph{2015 IEEE 82nd Vehicular Technology Conference (VTC2015-Fall)}, 2015,
  pp. 1--5.

\bibitem[Nikopour and Baligh(2013)]{6666156}
H.~Nikopour and H.~Baligh, ``{Sparse code multiple access},'' in \emph{2013
  IEEE 24th Annual International Symposium on Personal, Indoor, and Mobile
  Radio Communications (PIMRC)}, 2013, pp. 332--336.

\bibitem[Bockelmann et~al.(2018)Bockelmann, Pratas, Wunder, Saur, Navarro,
  Gregoratti, Vivier, De~Carvalho, Ji, Stefanovic, Popovski, Wang, Schellmann,
  Kosmatos, Demestichas, Raceala-Motoc, Jung, Stanczak, and Dekorsy]{8360103}
C.~Bockelmann, N.~K. Pratas, G.~Wunder, S.~Saur, M.~Navarro, D.~Gregoratti,
  G.~Vivier, E.~De~Carvalho, Y.~Ji, C.~Stefanovic, P.~Popovski, Q.~Wang,
  M.~Schellmann, E.~Kosmatos, P.~Demestichas, M.~Raceala-Motoc, P.~Jung,
  S.~Stanczak, and A.~Dekorsy, ``{Towards Massive Connectivity Support for
  Scalable mMTC Communications in 5G Networks},'' \emph{IEEE Access}, vol.~6,
  pp. 28\,969--28\,992, 2018.

\bibitem[Chen et~al.(2019)Chen, Sohrabi, and Yu]{8734871}
Z.~Chen, F.~Sohrabi, and W.~Yu, ``{Multi-Cell Sparse Activity Detection for
  Massive Random Access: Massive MIMO Versus Cooperative MIMO},'' \emph{IEEE
  Transactions on Wireless Communications}, vol.~18, no.~8, pp. 4060--4074,
  2019.

\bibitem[Bana et~al.(2019)Bana, {de Carvalho}, Soret, Abrão, Marinello,
  Larsson, and Popovski]{BANA2019100859}
\BIBentryALTinterwordspacing
A.-S. Bana, E.~{de Carvalho}, B.~Soret, T.~Abrão, J.~C. Marinello, E.~G.
  Larsson, and P.~Popovski, ``{Massive MIMO for Internet of Things (IoT)
  connectivity},'' \emph{Physical Communication}, vol.~37, p. 100859, 2019.
  [Online]. Available:
  \url{http://www.sciencedirect.com/science/article/pii/S1874490719303891}
\BIBentrySTDinterwordspacing

\bibitem[De~Carvalho et~al.(2017)De~Carvalho, Bjornson, Sorensen, Popovski, and
  Larsson]{7891796}
E.~De~Carvalho, E.~Bjornson, J.~H. Sorensen, P.~Popovski, and E.~G. Larsson,
  ``{Random Access Protocols for Massive MIMO},'' \emph{IEEE Communications
  Magazine}, vol.~55, no.~5, pp. 216--222, 2017.

\bibitem[Björnson et~al.(2017)Björnson, de~Carvalho, Sørensen, Larsson, and
  Popovski]{bjornson_random_2017-1}
\BIBentryALTinterwordspacing
E.~Björnson, E.~de~Carvalho, J.~H. Sørensen, E.~G. Larsson, and P.~Popovski,
  ``{A Random Access Protocol for Pilot Allocation in Crowded Massive MIMO
  Systems},'' \emph{IEEE Transactions on Wireless Communications}, vol.~16,
  no.~4, pp. 2220--2234, Apr. 2017, arXiv: 1604.04248. [Online]. Available:
  \url{http://arxiv.org/abs/1604.04248}
\BIBentrySTDinterwordspacing

\bibitem[Marinello and Abrão(2019)]{8886093}
J.~C. Marinello and T.~Abrão, ``{Collision Resolution Protocol via Soft
  Decision Stochastic Retransmission},'' in \emph{2019 IEEE Wireless
  Communications and Networking Conference (WCNC)}, 2019, pp. 1--6.

\bibitem[Marinello et~al.(2020)Marinello, Abrão, Souza, de~Carvalho, and
  Popovski]{8935438}
J.~C. Marinello, T.~Abrão, R.~D. Souza, E.~de~Carvalho, and P.~Popovski,
  ``{Achieving Fair Random Access Performance in Massive MIMO Crowded
  Machine-Type Networks},'' \emph{IEEE Wireless Communications Letters},
  vol.~9, no.~4, pp. 503--507, 2020.

\bibitem[Ganesan et~al.(2020{\natexlab{a}})Ganesan, Björnson, and
  Larsson]{9154288}
U.~K. Ganesan, E.~Björnson, and E.~G. Larsson, ``{An Algorithm for Grant-Free
  Random Access in Cell-Free Massive MIMO},'' in \emph{2020 IEEE 21st
  International Workshop on Signal Processing Advances in Wireless
  Communications (SPAWC)}, 2020, pp. 1--5.

\bibitem[Chen et~al.(2018)Chen, Sohrabi, and Yu]{8462577}
Z.~Chen, F.~Sohrabi, and W.~Yu, ``{Sparse Activity Detection for Massive
  Connectivity in Cellular Networks: Multi-Cell Cooperation Vs Large-Scale
  Antenna Arrays},'' in \emph{2018 IEEE International Conference on Acoustics,
  Speech and Signal Processing (ICASSP)}, 2018, pp. 6618--6622.

\bibitem[Liu et~al.(2018)Liu, Larsson, Yu, Popovski, Stefanovic, and
  de~Carvalho]{8454392}
L.~Liu, E.~G. Larsson, W.~Yu, P.~Popovski, C.~Stefanovic, and E.~de~Carvalho,
  ``{Sparse Signal Processing for Grant-Free Massive Connectivity: A Future
  Paradigm for Random Access Protocols in the Internet of Things},'' \emph{IEEE
  Signal Processing Magazine}, vol.~35, no.~5, pp. 88--99, 2018.

\bibitem[Ke et~al.(2020)Ke, Gao, Wu, Gao, and Schober]{8961111}
M.~Ke, Z.~Gao, Y.~Wu, X.~Gao, and R.~Schober, ``{Compressive Sensing-Based
  Adaptive Active User Detection and Channel Estimation: Massive Access Meets
  Massive MIMO},'' \emph{IEEE Transactions on Signal Processing}, vol.~68, pp.
  764--779, 2020.

\bibitem[Dong et~al.(2019)Dong, Shi, and Ding]{8683324}
J.~Dong, Y.~Shi, and Z.~Ding, ``{Sparse Blind Demixing for Low-latency Signal
  Recovery in Massive Iot Connectivity},'' in \emph{ICASSP 2019 - 2019 IEEE
  International Conference on Acoustics, Speech and Signal Processing
  (ICASSP)}, 2019, pp. 4764--4768.

\bibitem[Fengler et~al.(2019)Fengler, Haghighatshoar, Jung, and
  Caire]{FenglerAsilomar}
A.~Fengler, S.~Haghighatshoar, P.~Jung, and G.~Caire, ``{Grant-Free Massive
  Random Access With a Massive MIMO Receiver},'' in \emph{2019 53rd Asilomar
  Conference on Signals, Systems, and Computers}, 2019, pp. 23--30.

\bibitem[Fengler et~al.(2021)Fengler, Haghighatshoar, Jung, and
  Caire]{FenglerTIT}
------, ``{Non-Bayesian Activity Detection, Large-Scale Fading Coefficient
  Estimation, and Unsourced Random Access With a Massive MIMO Receiver},''
  \emph{IEEE Transactions on Information Theory}, vol.~67, no.~5, pp.
  2925--2951, 2021.

\bibitem[Ganesan et~al.(2020{\natexlab{b}})Ganesan, Bj\"{o}rnson, and
  Larsson]{GanesanSPAWC}
U.~K. Ganesan, E.~Bj\"{o}rnson, and E.~G. Larsson, ``{An Algorithm for
  Grant-Free Random Access in Cell-Free Massive MIMO},'' in \emph{2020 IEEE
  21st International Workshop on Signal Processing Advances in Wireless
  Communications (SPAWC)}, 2020, pp. 1--5.

\bibitem[Ganesan et~al.(2021)Ganesan, Björnson, and Larsson]{GanesanTCOM}
U.~K. Ganesan, E.~Björnson, and E.~G. Larsson, ``{Clustering Based Activity
  Detection Algorithms for Grant-Free Random Access in Cell-Free Massive
  MIMO},'' \emph{IEEE Transactions on Communications}, pp. 1--1, 2021.

\bibitem[Marzetta et~al.(2016)Marzetta, Larsson, Yang, and
  Ngo]{marzetta2016fundamentals}
T.~L. Marzetta, E.~G. Larsson, H.~Yang, and H.~Q. Ngo, \emph{{Fundamentals of
  Massive MIMO}}.\hskip 1em plus 0.5em minus 0.4em\relax Cambridge University
  Press, 2016.

\bibitem[Bj\"{o}rnson et~al.(2017)Bj\"{o}rnson, Hoydis, and
  Sanguinetti]{bjornsonBook}
\BIBentryALTinterwordspacing
E.~Bj\"{o}rnson, J.~Hoydis, and L.~Sanguinetti, ``{Massive MIMO Networks:
  Spectral, Energy, and Hardware Efficiency},'' \emph{Foundations and Trends in
  Signal Processing}, vol.~11, no. 3–4, p. 154–655, Nov. 2017. [Online].
  Available: \url{https://doi.org/10.1561/2000000093}
\BIBentrySTDinterwordspacing

\bibitem[Callebaut et~al.(2021{\natexlab{a}})Callebaut, Gunnarsson, Guevara,
  Johansson, Van Der~Perre, and Tufvesson]{callebaut2021experimental}
G.~Callebaut, S.~Gunnarsson, A.~P. Guevara, A.~J. Johansson, L.~Van Der~Perre,
  and F.~Tufvesson, ``Experimental exploration of unlicensed sub-ghz massive
  mimo for massive internet-of-things,'' pp. 2195--2204, 2021.

\bibitem[Petajajarvi et~al.(2015)Petajajarvi, Mikhaylov, Roivainen, Hanninen,
  and Pettissalo]{7377400}
J.~Petajajarvi, K.~Mikhaylov, A.~Roivainen, T.~Hanninen, and M.~Pettissalo,
  ``{On the coverage of LPWANs: range evaluation and channel attenuation model
  for LoRa technology},'' in \emph{2015 14th International Conference on ITS
  Telecommunications (ITST)}, 2015, pp. 55--59.

\bibitem[Callebaut et~al.(2021{\natexlab{b}})Callebaut, Leenders, Van~Mulders,
  Ottoy, De~Strycker, and Van~der Perre]{s21030913}
\BIBentryALTinterwordspacing
G.~Callebaut, G.~Leenders, J.~Van~Mulders, G.~Ottoy, L.~De~Strycker, and
  L.~Van~der Perre, ``{The Art of Designing Remote IoT Devices --- Technologies
  and Strategies for a Long Battery Life},'' \emph{Sensors}, vol.~21, no.~3,
  2021. [Online]. Available: \url{https://www.mdpi.com/1424-8220/21/3/913}
\BIBentrySTDinterwordspacing

\bibitem[Senel and Larsson(2018)]{senel2018grant}
K.~Senel and E.~G. Larsson, ``{Grant-free massive MTC-enabled massive MIMO: A
  compressive sensing approach},'' \emph{IEEE Transactions on Communications},
  vol.~66, no.~12, pp. 6164--6175, 2018.

\end{thebibliography}
